\documentclass[12pt]{article}
\usepackage{graphicx}
\usepackage{amsmath}

\newtheorem{theorem}{Theorem}

\newtheorem{corollary}[theorem]{Corollary}

\newtheorem{definition}[theorem]{Definition}

\newtheorem{lemma}[theorem]{Lemma}

\newtheorem{remark}[theorem]{Remark}

\input{tcilatex}

\begin{document}

\title{Future global in time einsteinian spacetimes with U(1) isometry group}
\author{Yvonne\ Choquet-Bruhat and Vincent Moncrief}
\maketitle

\begin{abstract}
We prove that spacetimes satisfying the vacuum Einstein equations on a
manifold of the form $\Sigma\times U(1)\times R$ where $\Sigma$ is a compact
surface of genus $G>1$ and where the Cauchy data is invariant with respect
to U(1) and sufficiently small exist for an infinite proper time in the
expanding direction.
\end{abstract}

\section{Introduction}

In this paper we prove a global in time existence theorem, in the expanding
direction, for a family of spatially compact vacuum spacetimes having
spacelike U(1) isometry groups. The 4- manifolds we consider have the form $%
V=M\times R$ where $M$ is an (orientable) circle bundle over a compact
higher genus surface $\Sigma$ and where the spacetime metric is assumed to
be invariant with respect to the natural action of U(1) along the bundle's
circle fibers. We reduce Einstein's equations, \`{a} la Kaluza- Klein, to a
system on the base $\Sigma\times R$ where it takes the form of the 2+1
dimensional Einstein equations coupled to a wave map matter source whose
target space is the hyperbolic plane. This wave map represents the true
gravitational wave degrees of freedom that have descended from 3+1
dimensions to appear as ''matter'' degrees of freedom in 2+1 dimensions. The
2+1 metric itself contributes only a finite number of additional,
Teichmuller parameter, degrees of freedom which couple to the wave map and
control the conformal geometry of $\Sigma.$

After the constraints have been solved and coordinate conditions imposed,
through a well defined elliptic system, nothing remains but the evolution
problem for the wave map / Teichmuller parameter system though the latter
has now become non local in the sense that the ''background'' metric in
which the wave map is propagating is now a non local functional of the wave
map itself given by the solution of the elliptic system mentioned above.
Thus even in the special ''polarized'' case which we concentrate on here, in
which the wave map reduces to a pure wave equation, this wave equation is
now both non linear and non local.

In addition to the simplifying assumption of polarization (which obliges us
here to treat only trivial bundles, $M=\Sigma\times S^{1}$) we shall need a
smallness condition on the initial data, an assumption that the genus of $%
\Sigma$ is greater than 1 and a restriction on the initial values allowed
for the Teichmuller parameters. It seems straightforward to remove each of
these restrictions except for the smallness condition on the initial data.
In particular we believe that the methods developed herein can be extended
to the treatment of non polarized solutions on non trivial bundles over
surfaces including the torus (but not $S^{2})$ with no restriction on the
initial values of the Teichmuller parameters. Some preliminary work in this
direction has already been carried out.

We do not know how to remove the small data restriction even in the
polarized case but conjecture that long time existence should hold for
arbitrary large data since the U(1) isometry assumption seems to suppress
the formation of black holes (note that U(1) is here essentially a
''translational'' and not a ''rotational'' symmetry since the existence of
an axis of rotation would destroy the bundle structure). Of couse there is
as yet no large data global existence result for smooth wave maps in 2+1
dimensions even on a given background so there is no immediate hope for such
a result in our still more non linear (and non local) problem but the
polarized case, though non linear and non local as well, seems more
promising. One knows how to control the Teichmuller parameters in pure 2+1
gravity and a wave equation on a given curved background offers no special
difficulty. But now the ''background metric is instead a funtional of the
evolving scalar field and one needs to control this along with the
Teichmuller parameters.

Serious progress on this problem would represent a ''quantum jump'' forward
in one's understanding of long time existence problems for Einstein's
equations since, up to now, the only large data global results require
simplifying assumptions that effectively reduce the number of spatial
dimensions to one (e.g., Gowdy models and their generalizations, plane
symmetric gravitational waves, spherically symmetric matter coupled to
gravity) or zero (e.g., Bianchi models, 2+1 gravity). We hope that this work
on small data global existence will lay the groundwork for such an eventual
quantum jump.

But why assume a Killing field if only small data results are aimed for in
the current project? A small data global existence result already exists
(Andersson -Moncrief, in preparation) for Einstein equations on different
3-manifolds of negative Yamabe class which makes no symmetry assumption
whatsoever. Shouldn't those methods be applicable to our problem in which
case the U(1) symmetry assumption could be removed. The answer to this
question is far from obvious for a somewhat subtle reason. In those cases
where small data global existence can be established the conformal geometry
of the spatial slices (which represents the propagating gravitational wave
degrees of freedom) is tending to a well behaved limit. Therefore the
various Sobolev ''constants'' (which are in fact functionals of the
geometry) which are needed in the associated energy estimates are tending to
well behaved limits as well. This simplifying feature is however missing in
the current problem since, during the course of our evolution, the conformal
geometry of the circle bundles under study is undergoing a kind of Cheeger-
Gromov collapse in which the circular fibers shrink to zero length and the
various related Sobolev ''constants'' may careen out of control making even
small data energy estimates much more difficult.

Of the various Thurston types of 3- geometries which compactify to negative
or zero Yamabe class manifolds \{$H^{3},H^{2}\times R,SL(2,R),Sol,Nil,R^{3})$
only the hyperbolics are immune from such degenerations and the remaining
(positive Yamabe class) Thurston types $\{S^{3},S^{2}\times R\}$ not only
are subject to Cheeger- Gromov type collapse but also to recollapse of the
actual physical geometry to ''big crunch'' singularities in the future
direction. By focusing on negative (or zero) Yamabe class manifolds which
exclude (due to Einstein's equations) the occurrence of maximal
hypersurfaces, that would signal the onset of recollapse to a big crunch, we
thereby concentrate on spacetimes that can expand indefinitely.

That such Cheeger- Gromov collapse can be expected in solutions of
Einstein's equations can be seen already in the basic compactified Bianchi
models wherein all the known solutions of negative Yamabe type except $H^{3}$
exhibit conformal collapse either along circular fibers \{$H^{2}\times
R,SL(2,R)\},$ or collapse along $T^{2}$ fibers \{$Sol$\}, or even total
collapse with non zero but bounded curvature of the Gromov ''almost flat''
variety $\{Nil\}$. The solutions we are considering here (of Thurston type $%
H^{2}\times R$ or eventually $SL(2,R)$ in non polarized generalizations)
extend results exhibiting such behavior to a large family of spatially
inhomogeneous spacetimes. We sidestep the extra complication of degenerating
Sobolev constants by imposing $U(1)$ symmetry and carrying out Kaluza Klein
reduction to work on a spatial manifold of hyperbolic type (though now a
2-dimensional one) for which, as we shall show, collapse and the
corresponding degeneracy of the needed Sobolev constants is suppressed.

The reason why we avoid the base $\Sigma=T^{2}$ is that the 2-tori
themselves tend to collapse under the Einstein flow whereas the higher genus
surfaces do not. On the other hand one can probably compute the explicit
dependence of the needed Sobolev constants on the Teichmuller parameters for
the torus and eventually exploit this to treat the Thurston cases \{$%
R^{3},Nil\}$ which compactify typically to trivial and non trivial $S^{1}-$%
bundles over $T^{2}.$ The $Sol$ case (which compactifies to $T^{2}$ bundles
over $S^{1})$ tends to collapse (as seen from the Bianchi models) the entire 
$T^{2}$ fibers. Thus to avoid degenerating Sobolev ''constants'' in this
case it seems necessary to impose a full $T^{2}=U(1)\times U(1)$ isometry
group and Kaluza Klein reduce to an $S^{1}$ spatial base manifold. This
leads to a certain nice generalization of the Gowdy models defined on the
''Sol - twisted torus'' but has effectively only one space dimension
remaining. We exclude the Thurston types \{$S^{2}\times R,S^{3}$\} which
correspond to trivial and non trivial $S^{1}$ bundles over $S^{2}$
respectively since they belong to the positive Yamabe class as we have
mentioned and should not exhibit infinite expansion but rather recollapse to
big crunch singularities.

The eight Thurston types are the basic building blocks from which other (and
conjucturally all) compact 3-manifolds can be built by glueing together
along so called incompressible 2-tori or (to obtain non prime manifolds)
along essential 2-spheres. Very little is known about the Einstein ''flow''
on such more general manifolds but it seems that a natural first step in
this direction may be made by studying the Einstein flow on the basic
building block manifolds themselves. This program seems tractable provided
that a $U(1)$ symmetry is imposed in the $H^{2}\times R,\widetilde{Sl(2,R)%
\text{ }}$ and perhaps $Nil$ and $R^{3}$ cases, and provided that a $%
U(1)\times U(1)$ symmetry is imposed in the $Sol$ case. No symmetries are
needed in the $H^{3}$ case due to the absence of Cheeger - Gromov collapse
but one can hope to remove the symmetry hypothesis in the other cases by
learning how to handle degenerating Sobolev ''constants''. In this respect
the $Nil$ and $R^{3}$ cases may provide some guidance since they seem to
require a treatment of degenerating Sobolev constants but only in the
setting of 2-dimensions (when $U(1)$ symmetry is imposed).

The basic methods we use involve the construction of higher order energies
to control the Sobolev norms of the scalar wave degrees of freedom combined
with an application of the ''Dirichlet energy'' function in Teichmuller
space to control the Teichmuller parameters degrees of freedom. A subtlety
is that the most obvious definition of wave equation (or, more generally,
wave map) energies does not lead to a well defined rate of decay so that
corrected energies must be introduced which exploit ''information'' about
the lowest eigenvalue of the spatial laplacian which enters into the wave
equation. Since the lowest eigenvalues vary with position in Teichmuller
space we find convenient to choose initial data such that, during the course
of the evolution, the lowest eigenvalue avoids a well known gap in the
spectrum for an arbitrary higher genus surface. If no eigenvalue drifts into
this gap (which we enforce by suitable restriction on the initial data) then
one can establish a universal rate of decay for the energies. If the lowest
eigenvalue \ drifts into this gap and remains there asymptotically then the
rate of decay of these energies will depend upon the asymptotic value of the
lowest eigenvalue and will no longer be universal. While it is
straightforward to modify the definitions of the corrected energies to take
this refinement into account we shall not do so here to avoid further
complication of an already involved analysis. An extension of the definition
of our corrected energies to the non polarized case and to the treatment of
non trivial $S^{1}$ bundles is also relatively straightforward but for
simplicity we shall not pursue that here either.

The sense in which our solutions are global in the expanding direction is
that they exhaust the maximal range allowed for the mean curvature function
on a manifold of negative Yamabe type, for which a zero mean curvature can
only be asymptotically approached. The normal trajectories to our space
slices all have an infinite proper time length. We do not attempt to prove
causal geodesic completeness but that would be straightforward to do given
the estimates we obtain.

Another question concerns the behavior of our solutions in the collapsing
direction. Since our energies are decaying in the expanding direction they
are growing in the collapsing direction and will eventually escape the
region in which we can control their behavior. In particular we cannot use
these arguments to show that our solutions extend to their conjectured
natural limit as the mean curvature function tends to $-\infty$. There is
another approach to the $U(1)$ problem however which, although local in
nature, can describe a large family of $U(1)$ - symmetric spacetimes by
convergent expansions about the big-bang singularities themselves. This
method, which is based on work by S. Kichenassamy and its extensions by A.
Rendall and J.\ Isenberg, can handle vacuum spacetimes that are ''velocity
dominated'' at their big-bang singularities. Work by J. Isenberg and one of
us (V.M) shows that the polarized vacuum solutions on $T^{3}\times R$ are
amenable to this analysis. In fact there two larger families of ''half -
polarized'' solutions that can also be rigorously treated and shown to have
velocity dominated singularities. By contrast the general (non polarized)
solution does not seem to be amenable to this kind of analysis and indeed
numerical work by B.\ Berger shows that such solutions should have
generically ''oscillatory'' rather than velocity dominated singularities.
The expansion methods which produce these solutions near their velocity
dominated singulariries are essentially local and should be readily
adaptable to other manifolds such as circle bundles over higher genus
surfaces. Thus one should be able to generate a large collection of initial
data sets for the problem dealt with in this papaer which treats the further
evolution globally in the expanding direction. Thus the machinery seems to
be at hand for treating a large family of $U(1)$ symmetric solutions from
their big - bang initial singularities to the limit of infinite
expansion.\bigskip

\section{Equations.}

The spacetime manifold $V$ is a principal fiber bundle with one dimensional
Lie group G and base $\Sigma\times R$, with $\Sigma$ a smooth 2 dimensional
manifold which we suppose here to be compact.

The spacetime metric is invariant under the action of G, the orbits are the
fibers of V and are supposed to be space like. We write it in the form 
\begin{equation*}
^{(4)}g=e^{-2\gamma\text{ }(3)}g+e^{2\gamma}(\theta)^{2},
\end{equation*}
where $\gamma$ is a scalar function and $^{(3)}g$ a lorentzian metric on $%
\Sigma\times R$ which reads: 
\begin{equation*}
^{(3)}g=-N^{2}dt^{2}+g_{ab}(dx^{a}+\nu^{a}dt)(dx^{b}+\nu^{b}dt)
\end{equation*}

\noindent N and $\nu$ are respectively the lapse and shift of $^{(3)}g,$%
while 
\begin{equation*}
g=g_{ab}dx^{a}dx^{b}
\end{equation*}

\noindent is a riemannian metric on $\Sigma$, depending on t.

The 1-form $\theta$ is a connection on the fiber bundle V, represented in
coordinates $(x^{3},x^{\alpha})$ adapted to the bundle structure by 
\begin{equation*}
\theta=dx^{3}+A_{\alpha}dx^{\alpha}
\end{equation*}
Note that $A$ is a locally defined 1-form on $\Sigma\times R.$

\subsection{Twist potential.}

The curvature of the connection locally represented by $A$ is a 2-form $A$
on $\Sigma\times R$, given by 
\begin{equation*}
F_{\alpha\beta}=(1/2)e^{-3\gamma}\eta_{\alpha\beta\lambda}E^{\lambda}
\end{equation*}
where E is an arbitrary closed 1-form if the equations $^{(4)}R_{\alpha3}$=$%
0 $ are satisfied. Hence if $\Sigma$ is compact 
\begin{equation*}
E=d\omega+H
\end{equation*}
where $\omega$ is a scalar function on V, called the twist potential, and $H 
$ a representative of the 1-cohomology class of $\Sigma\times R$, for
instance defined by a 1-form on $\Sigma$, harmonic for some given riemannian
metric m.

\subsection{Wave map equation.}

The fact that $F$ is a closed form together with the equation $%
^{(4)}R_{33}=0 $ imply (with the choice $H$=0) that the pair $%
u\equiv(\gamma,\omega)$ satisfies a wave map equation from ($\Sigma\times
R,^{(3)}g)$ into the hyperbolic 2-space, i.e. $R^{2}$ endowed with the
riemannian metric $2(d\gamma)^{2}+(1/2)e^{4\gamma}(d\omega)^{2}).$ This wave
map equation is a system of hyperbolic type when $^{(3)}g$ is a known
lorentzian metric.

\emph{In this article we will consider only the polarized case that is we
take }$\omega$ \emph{and H to be zero. }Some of the computations and partial
results hold however in the general case. It is why we keep the wave map
notation wherever possible, since we intend to extend our final result to
the general case in later work.

In the polarized case the wave map equation reduces to the wave equation for 
$\gamma$ in the metric $^{(3)}g.$

\subsection{3-dimensional Einstein equations}

When $^{(4)}R_{3\alpha}=0$ and $^{(4)}R_{33}=0$ the Einstein equations $%
^{(4)}R_{\alpha\beta}=0$ are equivalent to Einstein equations on the
3-manifold $\Sigma\times R$ for the metric $^{(3)}g$ with source the stress
energy tensor of the wave map: 
\begin{equation}
^{(3)}R_{\alpha\beta}=\partial_{\alpha}u.\partial_{\beta}u
\end{equation}
where a dot denotes a scalar product in the metric of the hyperbolic
2-space. We continue to use the same notation in the polarized case, that is
we set $\gamma=u$ and 
\begin{equation*}
\partial_{\alpha}u.\partial_{\beta}u\equiv2\partial_{\alpha}\gamma
\partial_{\beta}\gamma.
\end{equation*}

These Einstein equations decompose into

a. Constraints.

b. Equations for lapse and shift to be satisfied on each $\Sigma_{t}$. These
equations, as well as the constraints, are of elliptic type.

c. Evolution equations for the Teichmuller parameters, which are ordinary
differential equations.

\subsubsection{ Constraints on $\Sigma_{t}$.}

One denotes by $k$ the extrinsic curvature of $\Sigma_{t}$ as submanifold of
($\Sigma\times R,^{(3)}g);$ then, with $\nabla$ the covariant derivative in
the metric $g,$%
\begin{equation*}
k_{ab}\equiv(2N)^{-1}(-\partial_{t}g_{ab}+\nabla_{a}\nu_{b}+\nabla_{b}%
\nu_{a})
\end{equation*}
The equations (momentum constraint) 
\begin{equation}
^{(3)}R_{0a}\equiv N(-\nabla_{b}k_{a}^{b}+\partial_{a}\tau)=\partial
_{0}u.\partial_{a}u
\end{equation}
and (hamiltonian constraint, $^{(3)}S_{00}\equiv^{(3)}R_{00}+\frac{1}{2}%
N^{2} $ $^{(3)}R$) 
\begin{equation}
2N^{-2(3)}S_{00}\equiv R(g)-k_{b}^{a}k_{a}^{b}+\tau^{2}=N^{-2}\partial
_{0}u.\partial_{0}u+g^{ab}\partial_{a}u.\partial_{b}u
\end{equation}
do not contain second derivatives transversal to $\Sigma_{t}$ of g or u,
they are the constraints. To transform the constraints into an elliptic
system one uses the conformal method. We set 
\begin{equation*}
g_{ab}=e^{2\lambda}\sigma_{ab},
\end{equation*}
where $\sigma$ is a riemannian metric on $\Sigma$, depending on t, on which
we will comment later, and 
\begin{equation*}
k_{ab}=h_{ab}+\frac{1}{2}g_{ab}\tau
\end{equation*}
where $\tau$ is the $g$-trace of $k$, hence $h$ is traceless.

We denote by $D$ a covariant derivation in the metric $\sigma$. From now on,
unless otherwise specified, all operators are in the metric $\sigma$, and
indices are raised or lowered in this metric. We set 
\begin{equation*}
u^{\prime}=N^{-1}\partial_{0}u
\end{equation*}
with $\partial_{0}$ the Pfaff derivative of $u$, namely 
\begin{equation*}
\partial_{0}=\partial_{t}-\nu^{a}\partial_{a}\text{ with }\partial_{a}=\frac{%
\partial}{\partial x^{a}}
\end{equation*}
and 
\begin{equation*}
\overset{.}{u}=e^{2\lambda}u^{\prime}
\end{equation*}

The momentum constraint reads if $\tau$ is constant in space, a choice which
we will make, 
\begin{equation}
D_{b}h_{a}^{b}=L_{a},L_{a}\equiv-D_{a}u.\overset{.}{u}
\end{equation}
This is a linear equation for $h,$ independent of $\lambda.$ The general
solution is the sum of a transverse traceless tensor $h_{TT}\equiv q$ and a
conformal Lie derivative $r.$ Such tensors are $L^{2}-$orthogonal on $%
(\Sigma,\sigma).$

The hamiltonian constraint reads as the semilinear elliptic equation in $%
\lambda:$ 
\begin{equation}
\Delta\lambda=f(x,\lambda)\equiv p_{1}e^{2\lambda}-p_{2}e^{-2\lambda}+p_{3},
\end{equation}
with 
\begin{equation*}
p_{1}\equiv\frac{1}{4}\tau^{2},p_{2}\equiv\frac{1}{2}(\mid\overset{.}{u}%
\mid^{2}+\mid h\mid^{2}),p_{3}\equiv\frac{1}{2}(R(\sigma)-|Du|^{2})
\end{equation*}

\subsubsection{ Equations for lapse and shift.}

The lapse and shift are gauge parameters for which we obtain elliptic
equations on each $\Sigma_{t}$ as follows.

We impose that the $\Sigma_{t}^{\prime}s$ have constant (in space) mean
curvature, namely that $\tau$ is a given negative increasing function of $t$%
. The lapse $N$ satisfies then the linear elliptic equation 
\begin{equation*}
\Delta N-\alpha N=-e^{2\lambda}\partial_{t}\tau
\end{equation*}
with 
\begin{equation*}
\alpha\equiv e^{-2\lambda}(\mid h\mid^{2}+\mid\overset{.}{u}\mid^{2})+\frac
{1}{2}e^{2\lambda}\tau^{2}
\end{equation*}
The equation to be satisfied by the shift $\nu$ results from the knowledge
of $\sigma_{t}$. Indeed the definition of $k$ implies that $\nu$ satisfies a
linear differential equation with an operator $L$, the conformal Lie
derivative, which we first write in the metric $g$: 
\begin{equation*}
(L_{g}\nu)_{ab}\equiv\nabla_{a}\nu_{b}+\nabla_{b}\nu_{a}-g_{ab}\nabla_{c}%
\nu^{c}=\phi_{ab}
\end{equation*}
with 
\begin{equation*}
\phi_{ab}\equiv2Nh_{ab}+\partial_{t}g_{ab}-\frac{1}{2}g_{ab}g^{cd}\partial
_{t}g_{cd}
\end{equation*}
then in the metric $\sigma$%
\begin{equation*}
(L_{\sigma}n)_{ab}\equiv D_{a}n_{b}+D_{b}n_{a}-\sigma_{ab}D_{c}n^{c}=f_{ab}%
\text{ with }n_{a}\equiv\nu_{a}e^{-2\lambda}
\end{equation*}
where 
\begin{equation*}
f_{ab}\equiv2Ne^{-2\lambda}h_{ab}+\partial_{t}\sigma_{ab}-\frac{1}{2}%
\sigma_{ab}\sigma^{cd}\partial_{t}\sigma_{cd}
\end{equation*}

The kernel of the dual of $L$ is the space of transverse traceless symmetric
2-tensors, i.e. symmetric 2-tensors $T$ such that

\begin{equation}
g^{ab}T_{ab}=0,\text{ \ \ }\nabla^{a}T_{ab}=0.
\end{equation}
These tensors are usually called TT tensors. The spaces of TT tensors are
the same for two conformal metrics.

\subsubsection{ Teichmuller parameters.}

On a compact 2-dimensional manifold of genus G $\geq2$ the space $T_{eich}$
of conformally inequivalent riemannian metrics, called Teichmuller space,
can be identified (cf. Fisher and Tromba) with $M_{-1}/D_{0}$, the quotient
of the space of metrics with scalar curvature $-1$ by the group of
diffeomorphisms homotopic to the identity. $M_{-1}\mathcal{\rightarrow}%
T_{eich}$ is a trivial fiber bundle whose base can be endowed with the
structure of the manifold $R^{n}$, with $n=6G-6$.

We impose to the metric $\sigma_{t}$ to be in some chosen cross section $%
Q\rightarrow\psi(Q)$ of the above fiber bundle. Let $Q^{I},I=1,...,n$ be
coordinates in $T_{eich}$, then $\partial\psi/\partial Q^{I}$ is a known
tangent vector to $M_{-1}$ at $\psi(Q)$, that is a traceless symmetric
2-tensor field on $\Sigma,$ sum of a transverse traceless tensor field $%
X_{I}(Q)$ and of the Lie derivative of a vector field on the manifold $%
(\Sigma,\psi(Q))$. The tensor fields $X_{I}(Q),I=1,...n$ span the space of
transverse traceless tensor fields on $(\Sigma,\psi(Q)).$ The matrix with
elements 
\begin{equation*}
\int_{\Sigma}X_{I}^{ab}X_{Jab}\mu_{\psi(Q)}
\end{equation*}
is invertible.

\begin{lemma}
If we impose to the metric $\sigma_{t}$ to lie in the chosen cross section,
i.e. $\sigma_{t}\equiv\psi(Q(t))$, the solvability condition for the shift
equation determines $dQ^{I}/dt$ in terms of $h_{t}$.
\end{lemma}

Proof.

The time derivative of $\sigma$ is given by

\begin{equation*}
\partial_{t}\sigma=(dQ^{I}/dt)\partial\psi/\partial Q^{I}
\end{equation*}
hence it is of the form 
\begin{equation*}
\partial_{t}\sigma_{ab}=\frac{dQ^{I}}{dt}X_{Iab}+C_{ab}
\end{equation*}
where $C$ is a Lie derivative, $L^{2}$ orthogonal to TT tensors.

The shift equation on $\Sigma_{t}$ is solvable if and only if its right hand
side $f$ is $L^{2}$ orthogonal to TT tensors, i.e. to each tensor field $%
X_{I}$. Theses conditions read 
\begin{equation*}
\int_{\Sigma_{t}}f_{ab}X_{Jab}\mu_{\sigma_{t}}=0
\end{equation*}
We have seen that $h$ is the sum of a tensor $r$ which is in the range of
the conformal Killing operator, hence $L^{2}$ orthogonal to TT tensors, and
a TT tensor. This last tensor can be written with the use of the basis $%
X_{I} $ of such tensors, the coefficients $P^{I}$ depending only on $t$: 
\begin{equation*}
h_{ab}^{TT}=P^{I}(t)X_{I,ab}
\end{equation*}
The orthogonality conditions read, using the fact that the transverse
tensors $X_{I}$ are orthogonal to Lie derivatives and are traceless: 
\begin{equation*}
\int_{\Sigma_{t}}[2Ne^{-2%
\lambda}(r_{ab}+P^{I}X_{I,ab})+(dQ^{I}/dt)X_{I,ab}]X_{j}^{ab}\mu_{\sigma}=0
\end{equation*}
The tangent vector $dQ^{I}/dt$ to the curve $t\rightarrow Q(t)$ and the
tangent vector $P^{I}(t)$ to $T_{eich}$ are therefore linked by the linear
system 
\begin{equation*}
X_{IJ}\frac{dQ^{I}}{dt}+Y_{IJ}P^{I}+Z_{J}=0
\end{equation*}
with 
\begin{equation*}
X_{IJ}\equiv\int_{\Sigma_{t}}X_{I}^{ab}X_{J,ab}\mu_{\sigma_{t}}
\end{equation*}

\begin{equation*}
Y_{IJ}\equiv\int_{\Sigma_{t}}2Ne^{-2\lambda}X_{I}^{ab}X_{J,ab}\mu_{%
\sigma_{t}}
\end{equation*}
\begin{equation*}
Z_{J}\equiv\int_{\Sigma_{t}}2Ne^{-2\lambda}r_{ab}X_{J}^{ab}\mu_{\sigma_{t}}
\end{equation*}

We will now construct an ordinary differential system for the evolution of
the $Q^{I}$ and $P^{I}$ by considering the as yet non solved 3-dimensional
Einstein equations 
\begin{equation*}
^{(3)}R_{ab}=\rho_{ab}\equiv\partial_{a}u.\partial_{b}u
\end{equation*}

\begin{lemma}
The constraint equations together with the lapse and the wave map equations
imply that $N(^{(3)}R_{ab}-\rho_{ab})$ with $\rho_{ab}\equiv\partial
_{a}u.\partial_{b}u$ is a transverse traceless tensor on each $\Sigma_{t}$.
\end{lemma}

Proof. 1. The equations

\begin{equation}
^{(3)}S_{00}=T_{00}
\end{equation}
and 
\begin{equation}
^{(3)}R_{00}=\rho_{00}
\end{equation}
imply 
\begin{equation}
^{(3)}R=\rho
\end{equation}
since 
\begin{equation}
^{(3)}S_{00}-T_{00}\equiv^{(3)}R_{00}-\frac{1}{2}g_{00}^{(3)}R-(\rho _{00}-%
\frac{1}{2}g_{00}\text{ }\rho)
\end{equation}
hence 
\begin{equation*}
^{(3)}R^{ab}-\rho^{ab}=^{(3)}S^{ab}-T^{ab}
\end{equation*}
The equations 2 and 3 imply

\begin{equation*}
g^{ab}(^{(3)}R_{ab}-\rho_{ab})=0
\end{equation*}
On the other hand the Bianchi identity in the 3-metric g gives 
\begin{equation*}
^{(3)}\nabla_{\alpha}(^{(3)}S^{\alpha b}-T^{\alpha b})=0.
\end{equation*}
An elementary calculus using the connexion coefficients of $^{(3)}g$ and $g$
shows that, due to equations previously satisfied, this equation reduces to
the following divergence in the metric $g$: 
\begin{equation*}
\nabla_{a}[N(^{(3)}R^{ab}-\rho^{ab})]=0
\end{equation*}
The tensor $N$($^{(3)}R_{ab}-\rho_{ab})$ is therefore a traceless and
transverse tensor on ($\Sigma,g)$, and hence also on ($\Sigma$,$\sigma)$, by
conformal invariance of this property for symmetric 2-covariant tensors.

We deduce from this lemma that a necessary and sufficient condition for the
previous equations to imply $^{(3)}R_{ab}-\rho_{ab}=0$ is that the tensor $%
N(^{(3)}R_{ab}-\rho_{ab})$ be $L^{2}$ orthogonal to transverse traceless
tensors on ($\Sigma_{t},\sigma_{t})$, i.e. to each of the TT tensors $X_{I} $
defined above through the cross section $\psi$ where we choose $\sigma_{t}$
, that is 
\begin{equation*}
\int_{\Sigma_{t}}N(^{(3)}R_{ab}-\rho_{ab})X_{J}^{ab}\mu_{\sigma_{t}}=0,\text{
for }J=1,2,...6G-6
\end{equation*}
We recall that

\begin{equation*}
^{(3)}R_{ab}\equiv R_{ab}-N^{-1}\overset{\_}{\partial }%
_{0}k_{ab}-2k_{ac}k_{b}^{c}+\tau k_{ab}-N^{-1}\nabla _{a}\partial _{b}N
\end{equation*}
with 
\begin{equation*}
k_{ab}\equiv P^{I}X_{I,ab}+r_{ab}+\frac{1}{2}g_{ab}\tau
\end{equation*}
and $\overset{\_}{\partial }_{0}$ is an operator on time dependent space
tensors (cf. C-B and York) defined by, with $\mathcal{L}_{\nu }$ the Lie
derivative in the direction of $\nu ,$%
\begin{equation*}
\overset{\_}{\partial }_{0}\equiv \partial _{t}-\mathcal{L}_{\nu }
\end{equation*}
We thus obtain an ordinary differential system of the form 
\begin{equation*}
X_{IJ}\frac{dP^{I}}{dt}+\Phi _{J}(P,\frac{dQ}{dt})=0
\end{equation*}
where $\Phi $ is a polynomial of degree 2 in $P$ and $dQ/dt$ with
coefficients depending smoothly on $Q$ and directly but continuously on $t$
through the other unknown, namely: 
\begin{equation*}
\Phi _{J}\equiv A_{JIK}P^{I}P^{K}+B_{JIK}P^{I}\frac{dQ^{K}}{dt}%
+C_{JI}P^{I}+D_{J}
\end{equation*}
with 
\begin{equation*}
A_{JIK}\equiv \int_{\Sigma _{t}}2Ne^{2\lambda
}X_{I,a}^{c}X_{K,bc}X_{J}^{ab}\mu _{\sigma _{t}}
\end{equation*}
\begin{equation*}
B_{JIK}\equiv \int_{\Sigma _{t}}\frac{\partial X_{I,ab}}{\partial Q^{K}}%
X_{J}^{ab}\mu _{\sigma _{t}}
\end{equation*}
\begin{equation*}
C_{JI}\equiv \int_{\Sigma _{t}}[(-\mathcal{L}_{\nu
}X_{I})_{ab}+4Ne^{-2\lambda }r_{b}^{c}X_{I,ac}-\tau NX_{I,ab}]X_{J}^{ab}\mu
_{\sigma _{t}}
\end{equation*}
and, using integration by parts and the transverse property of the $X_{I%
\text{ }}$to eliminate second derivatives of $N$ (recall that $\nabla
_{a}\partial _{b}N\equiv D_{a}\partial _{b}N-2\partial _{a}\lambda \partial
_{b}N)$ 
\begin{equation*}
D_{J}\equiv \int_{\Sigma _{t}}(-\overset{\_}{\partial }_{0}r_{ab}-2Ne^{-2%
\lambda }r_{ac}r_{b}^{c}+\tau Nr_{ab}+2\partial _{a}\lambda \partial
_{b}N-\partial _{a}u.\partial _{b}u)X_{J}^{ab}\mu _{\sigma _{t}}.
\end{equation*}

\section{Homogeneous solution.}

\begin{theorem}
A particular solution, obtained by taking for u a constant wave map and for
h the zero tensor, is given by: 
\begin{equation*}
^{(4)}g=-4dt^{2}+2t^{2}\sigma_{ab}dx^{a}dx^{b}+\theta^{2}
\end{equation*}
with $\sigma$ a metric on $\Sigma$ independent of t and of scalar curvature $%
-1$, and $\theta$ a flat connexion 1-form on the bundle.
\end{theorem}

Proof. The wave map equation is satisfied by any constant map. Such a map
has zero stress energy tensor. The momentum constraint is then satisfied by
h = 0, hence 
\begin{equation*}
k_{ab}=\frac{1}{2}\tau g_{ab}
\end{equation*}
The hamiltonian constraint is satisfied by a constant in space $\lambda$
given if $R(\sigma)=-1$ by 
\begin{equation*}
e^{2\lambda}=\frac{2}{\tau^{2}}
\end{equation*}
the shift equation is then satisfied by $\nu=0$ and the lapse equation by 
\begin{equation*}
N=2
\end{equation*}
A straightforward computation shows that $^{(3)}R_{ab}=0.$ All the equations 
$Ricci(^{(4)}g)=0$ are satisfied.

\begin{remark}
The hypothesis imply that the bundle M$\rightarrow\Sigma$ is a trivial
bundle.
\end{remark}

\section{Local existence theorem.}

\subsection{Cauchy problem.}

The unknowns which permit the reconstruction of the spacetime metric in the
gauge $\tau\equiv\tau(t)$, given some smooth cross section $Q\rightarrow
\psi(Q)$ of Teichmuller space $T_{eich}$, are on the one hand $u=\gamma$
satisfying the wave equation in the metric $^{(3)}g$, on the other hand $%
\lambda,$ $N$ and $\nu$, which satisfy elliptic equations on each $\Sigma
_{t}$, and also a curve $Q(t)$ in $T_{eich}$ which determines the metric $%
\sigma_{t}\equiv\psi(Q(t))$ on $\Sigma_{t}.$ An intermediate unknown is the
traceless tensor $h$ which splits into a transverse part and a conformal Lie
derivative of $\sigma_{t}$ in the direction of a vector $Y$ which satisfies
also an elliptic system on $\Sigma_{t}.$ The tranverse part is determined
through a field of tangent vectors to $T_{eich}$ at the points of $Q(t).$

\begin{definition}
The Cauchy data on $\Sigma_{t_{0}}$ denoted $\Sigma_{0}$ are:
\end{definition}

1. A $C^{\infty}$ riemannian metric $\sigma_{0}$ which projects onto a point 
$Q(t_{0})$ of $T_{eich}$ and a $C^{\infty}$ tensor $q_{0}$ transverse and
traceless in the metric $\sigma_{0}.$

2. Cauchy data for $u$ and $\overset{.}{u}$ on $\Sigma_{0}$, i.e. 
\begin{equation*}
u(t_{0},.)=u_{0}\in H_{2},\overset{.}{u}(t_{0},.)=\overset{.}{u}_{0}\in H_{1}
\end{equation*}
where $H_{s}$ is the usual Sobolev space on $(\Sigma,\sigma_{0}).$

\subsection{Functional spaces.}

\begin{definition}
Let $\sigma_{t}$ be a curve of $C^{\infty}$ riemannian metrics on $\Sigma,$
uniformly equivalent to the metric $\sigma_{0}$ for t$\in\lbrack t_{0},T]$
and $C^{1}$ in t. Such metrics are called regular for $t\in\lbrack t_{0},T]$
\end{definition}

1. The spaces W$_{s}^{p}(t)$ are the usual Sobolev spaces of tensor fields
on the riemannian manifold $(\Sigma,\sigma_{t}).$

By the hypothesis on $\sigma_{t}$ the norms in $W_{s}^{p}(t)$ are uniformly
equivalent for $t\in\lbrack t_{0,},T]$ to the norm in $W_{s}^{p}(t_{0}).$ We
set $W_{s}^{2}(t)=H_{s}(t).$ When working on one slice $\Sigma_{t}$ we will
often omit reference to the $t$ dependence of the norm.

2. The spaces $E_{s}^{p}(T)$ are the Banach spaces of t dependent tensor
fields $f$ on $\Sigma$ 
\begin{equation*}
E_{s}^{p}(T)\equiv C^{0}([t_{0},T],W_{s}^{p})\cap
C^{1}([t_{0},T],W_{s-1}^{p}).
\end{equation*}
with norm 
\begin{equation*}
||f||_{E_{s}^{p}(T)}=Sup_{t_{0}\leq t\leq
T}(||f||_{W_{s}^{p}(t)}+||\partial_{t}f||_{W_{s-1}^{p}(t)}).
\end{equation*}
\bigskip We set $E_{s}^{2}(T)=E_{s}(T).$

We will proceed in two steps:

Case a. $Du_{0},\overset{.}{u}_{0}\in H_{2}$

Case b. $Du_{0},\overset{.}{u}_{0}\in H_{1}$

We will need the following lemma (we set \ $E_{s}=E_{s}^{2})$

\begin{lemma}
Let $\sigma_{t}$ be a regular metric on $\Sigma\times\lbrack t_{0},T]$ then
\end{lemma}

a. If $Du,\overset{.}{u}\in E_{2}(T)$ then $Du.Du,Du.\overset{.}{u},\overset{%
.}{u}.\overset{.}{u}\in E_{2}(T)$,

b. If $Du,\overset{.}{u}\in E_{1}(T)$ then $Du.Du,Du.\overset{.}{u},\overset{%
.}{u}.\overset{.}{u}\in E_{1}^{p}(T)\cap E_{0}^{q}(T)$, $1\leq p<2,$ $1\leq
q<\infty.$

Proof. a. Since in dimension 2 the space $H_{2}$ is an algebra one has 
\begin{equation*}
Du.Du,\overset{.}{u}\in C^{0}([t_{0},T],H_{2})
\end{equation*}
On the other hand we have 
\begin{equation*}
|\partial_{t}(Du.Du)|=2|\partial_{t}Du.Du|\leq2|\partial_{t}Du||Du|
\end{equation*}
hence by multiplication properties of Sobolev spaces 
\begin{equation*}
\partial_{t}(Du.Du)\in C^{0}([t_{0},T],H_{1})
\end{equation*}

b. If $Du\in E_{1}$ then $Du\in E_{0}^{q}\equiv C^{0}([t_{0},T],L^{q})$ ,
for all $q<\infty$ by the standard Sobolev embedding theorem, and so does $%
Du.Du$.

\noindent We have 
\begin{equation*}
|D(Du.Du)|=2|D^{2}u.Du|\leq2|D^{2}u||Du|
\end{equation*}
hence $D(Du.Du)\in E_{0}^{p}$ for all $1\leq p<2$ since $D^{2}u\in E_{0}$
and $Du\in E_{0}^{q},1\leq q<\infty.$\bigskip

\noindent An analogous proof gives the result for the other products.

Using again $|\partial_{t}(Du.Du)||\leq2|\partial_{t}Du||Du|$ we obtain $%
\partial_{t}(Du.Du)\in E_{0}^{p}$ for 1$\leq p<2$ since we have by
definition $\partial_{t}Du\in E_{0}^{2}$. Analogous reasonning with $%
\overset{.}{u}$ completes the proof.

\subsection{Resolution of the elliptic equations for given Q(t), P(t) and u}

We have supposed chosen a smooth cross section $Q\rightarrow\psi(Q)$ of $%
M_{-1}$ over the Teichmuller space $T_{eich}$. We suppose given a $C^{1}$
curve $t\rightarrow Q(t)$ contained when $t\in\lbrack t_{0},T]$ in a compact
subset of $T_{eich},$ and a continuous set of tangent vectors $P$ to $%
T_{eich}$ at points of this curve. We are then given by lift to $M_{-1}$ a
regular metric $\sigma_{t}$ for $t\in\lbrack t_{0},T]$, with scalar
curvature -1, together with a smooth symmetric 2-tensor $h_{t}^{TT}\equiv
q_{t}$ transverse and traceless in the metric $\sigma_{t\text{ }}$ and
depending continuously on $t$.

\subsubsection{Determination of h.}

We have set 
\begin{equation*}
h_{ab}=q_{ab}+r_{ab}
\end{equation*}
where q and r are traceless q is transverse and r is a conformal Lie
derivative, i.e. 
\begin{equation*}
D_{a}q_{b}^{a}=0\text{ and }q_{a}^{a}=0
\end{equation*}

\begin{equation*}
r_{ab}=D_{a}Y_{b}+D_{b}Y_{a}-\sigma_{ab}D_{c}Y^{c}
\end{equation*}

\subparagraph{Determination of q.}

The traceless transverse tensor q on ($\Sigma_{t},\sigma_{t})$ is deduced by
lifting its given projection onto the tangent space to Teichmuller space at
the point $Q(t).$ It is smooth and depends continuously on $t\in\lbrack
t_{0,}T].$ Let us denote by $X_{I}(Q),I=1,...,6G-6,$ a basis of traceless
transverse tensor fields on $(\Sigma,\psi(Q))$ then 
\begin{equation*}
q_{t}=X_{I}(Q(t))P^{I}(t)
\end{equation*}

\subparagraph{Determination of r.}

The vector Y satisfies on each $\Sigma_{t}$ the elliptic system with zero
kernel ( in accordance with the fact that ($\Sigma,\sigma)$ does not admit
conformal Killing fields when $R(\sigma)<0),$%
\begin{equation*}
D^{a}r_{ab}\equiv D^{a}D_{a}Y_{b}+\frac{1}{2}R(\sigma)Y_{b}=L_{b}\equiv
-D_{b}u.\overset{.}{u}
\end{equation*}

\subparagraph{Case a., $L\in E_{2}(T).$}

It results from elliptic theory that the system satisfied by Y has for each $%
t\in\lbrack0,T]$ one and only one solution in $H_{4}(t)$ and there exists a
constant depending only on $\sigma_{t}$ such that 
\begin{equation*}
||r_{t}||_{H_{3}(t)}\leq C_{\sigma_{t}}||Du.\overset{.}{u}||_{H_{2}(t)}
\end{equation*}
The constant $C_{\sigma_{t}}$ is invariant under diffeomorphism acting on $%
\sigma_{t}$, that is it depends only on its projection on the Teichmuller
space of $\Sigma$, hence is uniformly bounded under the hypothesis made on $%
\sigma_{t}$. We denote by $M_{\sigma,T}$ such a constant.

We have since the norms $W_{s}^{p}(t)$ and $W_{s}^{p}$ are uniformly
equivalent

\begin{equation*}
||r_{t}||_{H_{3}}\leq M_{\sigma,T}||(Du.\overset{.}{u})_{t}||_{H_{2}}
\end{equation*}
\bigskip

Derivations with respect to t of the equation for Y show that for a regular $%
\sigma_{t}$ we have 
\begin{equation*}
r\in E_{3}(T),||r||_{E_{3}(T)}\leq M_{\sigma,T}||Du||_{E_{2}(T)}\times||%
\overset{.}{u}||_{E_{2}(T)}
\end{equation*}

\subparagraph{Case b. $L\in E_{1}^{p}(T).$}

The system for Y has one and only one solution in $W_{3}^{p}(t)$ for each t$%
\in\lbrack t_{0},T]$, then r$_{t}$ is in $W_{2}^{p}(t)$ and there exists a
constant $C_{\sigma_{t}}$ such that

\begin{equation*}
||r_{t}||_{W_{2}^{p}(t)}\leq C_{\sigma_{t}}||(Du.\overset{.}{u}%
)_{t}||_{W_{1}^{p}(t)}
\end{equation*}

One proves also that $\partial_{t}r\in W_{1}^{p}(t)$ hence $r\in
E_{2}^{p}(T) $\ and there exists a constant $M_{\sigma,T}$ such that 
\begin{equation*}
||r||_{E_{2}^{p}(T)}\leq M_{\sigma,T}||Du||_{E_{1}(T)}\times||\overset{.}{u}%
||_{E_{1}(T)}
\end{equation*}

\subsubsection{Case of initial values.}

On the initial manifold $\Sigma_{t_{0}}$ we have given $q_{0}\in C^{\infty} $%
, and $r_{0}$ satisfies the inequality (we abbreviate to $%
\parallel.\parallel $ the $L^{2}$ norm on $(\Sigma,\sigma_{0}))$%
\begin{equation*}
\parallel r_{0}\parallel\leq C_{\sigma_{0}}\parallel Du_{0}.\overset{.}{u}%
_{0}\parallel
\end{equation*}
hence $h_{0}$ is small in $L^{2}$ norm if it is so of $q_{0}$ while $Du_{0}$
and $\overset{.}{u}_{0}$ are small in $H_{1}$ norm.

\subsubsection{Determination of the conformal factor $\protect\lambda$.}

On each $\Sigma_{t}$ the conformal factor $\lambda_{t}$ satisfies the
equation, with $\Delta\equiv\Delta_{\sigma_{t}}$ the laplacian in the metric 
$\sigma_{t}$ (we omit the writing of $t$ to simplify the notation) 
\begin{equation*}
\Delta\lambda=f(\lambda)\equiv p_{1}e^{2\lambda}-p_{2}e^{-2\lambda}+p_{3}
\end{equation*}
where the coefficients p are given by, with $R(\sigma)=-1,$ 
\begin{equation*}
p_{1}=\frac{1}{4}\tau^{2},p_{2}=\frac{1}{2}(\mid h\mid^{2}+\mid\dot{u}\mid
^{2}),p_{3}=\frac{1}{2}(R(\sigma)-\mid Du\mid^{2})
\end{equation*}

\subparagraph{Case a.}

We suppose that the coefficients p are given functions in $E_{2}(T).$ This
hypothesis is consistent with $Du,\dot{u}\in E_{2}(T)$ and $h\in E_{2}(T)$.

We know from elliptic theory that the semi linear elliptic equation for $%
\lambda$ on $(\Sigma_{t},\sigma_{t})$ admits a solution in $H_{4}(t)$, which
is included in $C^{2}$, if it admits a subsolution $\lambda_{-}$ and a
supersolution $\lambda_{+}$, i.e. $C^{2}$ functions such that

\begin{equation*}
\Delta\lambda_{+}\leq f(\lambda_{+}),\text{ \ \ and \ \ }\Delta\lambda_{-}%
\geq f(\lambda_{-})\text{, \ \ \ \ }\lambda_{-}\leq\lambda_{+}
\end{equation*}

We construct sub and super solutions as follows.

We define the number $\omega$ to be the real root of the equation 
\begin{equation*}
P_{1}e^{2\omega}-P_{2}e^{-2\omega}+P_{3}=0
\end{equation*}
where the P's are the integrals of the p's on $\Sigma_{t}$

By the Gauss Bonnet theorem the volume of $(\Sigma_{t},\sigma_{t})$ is a
constant if $R(\sigma_{t})$, is constant. We have here $R(\sigma_{t})=-1,$
hence: 
\begin{equation*}
V_{\sigma}\equiv\int_{\Sigma_{t}}\mu_{\sigma}=-\int_{\Sigma_{t}}R(\sigma
)\mu_{\sigma}=-4\pi\chi
\end{equation*}
We find: 
\begin{equation*}
P_{2}=\frac{1}{2}(\parallel h\parallel^{2}+\parallel\dot{u}%
\parallel^{2})\geq0
\end{equation*}
\begin{equation*}
P_{1}=\frac{1}{4}V_{\sigma}\tau^{2}>0,P_{3}=-\frac{1}{2}(V_{\sigma}+%
\parallel Du\parallel^{2})<0
\end{equation*}
hence $e^{2\omega}$ exists, is unique and satisfies 
\begin{equation*}
e^{2\omega}\geq2\tau^{-2}.
\end{equation*}
We define $v\in H_{4}$ as the solution with mean value zero on $\Sigma_{t}$
of the linear equation 
\begin{equation*}
\Delta v=f(\omega)\equiv p_{1}e^{2\omega}-p_{2}e^{-2\omega}+p_{3}
\end{equation*}
Such a solution exists and is unique, because $f(\omega)$ has mean value
zero on $\Sigma_{t}$

\begin{lemma}
. The functions $\lambda_{+}=\omega+v-min_{\Sigma}v$ and $%
\lambda_{-}=\omega+v-Max_{\Sigma}v$ are respectively a super and sub
solution of the equation for $\lambda$.
\end{lemma}

Proof . We have $\lambda_{+}\geq\omega$ and $\lambda_{-}\leq\omega$.hence $%
f(\lambda_{+})\geq f(\omega)\geq f(\lambda_{-})$, since $f$ is an increasing
function of $\lambda,$ while $\Delta\lambda_{+}=\Delta\lambda_{-}=\Delta
v=f(\omega)$.

The solution $\lambda$ $\in H_{4}$ thus obtained for each $t\in\lbrack
t_{0},T]$ is unique, due to the monotony of the function $f$. Its $H_{4}$
norm depends continuously on $t$. Derivation with respect to $t$ of the
equation satisfied by $\lambda$ shows that $\partial_{t}\lambda\in
C^{0}([t_{0},T],H_{3})$ .We have proved:

\begin{theorem}
The equation for $\lambda$ has one and only one solution $\lambda\in
E_{4}(T) $ under the hypothesis a (where $p_{i}\in E_{2}(T)).$
\end{theorem}

\subparagraph{Case b.}

\begin{theorem}
The equation for $\lambda$ has one and only one solution $\lambda\in
E_{3}^{p}(T)$ under the hypothesis b (where $p_{i}\in E_{1}^{p}(T)).$
\end{theorem}

Proof. Consider a Cauchy sequence of functions $p_{2}^{(n)}\geq0,p_{3}^{(n)}+%
\frac{1}{2}\leq0,$ both in $E_{2}(T)$, converging in $E_{1}^{p}(T)$ to
functions $p_{2},p_{3}+\frac{1}{2}.$ For each n there is a solution $%
\lambda_{(n)}\in E_{4}(T)$ of the conformal factor equation. The difference $%
\lambda_{(n)}-\lambda_{(m)}$ satisfies the equation 
\begin{equation*}
\Delta(\lambda_{(n)}-\lambda_{(m)})=p_{1}(e^{2\lambda_{(n)}}-e^{2\lambda
_{(m)}})-p_{2}^{(n)}(e^{-2\lambda_{(n)}}-e^{-2\lambda_{(m)}})
\end{equation*}
\begin{equation*}
+(p_{2}^{(m)}-p_{2}^{(n)})e^{-2\lambda_{(m)}}+p_{3}^{(n)}-p_{3}^{(m)}
\end{equation*}
Applying elementary calculus inequalities to the estimate of $%
(a-b)^{-1}(e^{a}-e^{b})$ and $(a-b)^{-1}(e^{-a}-e^{-b})$ one obtains a well
posed\ linear elliptic equation for $\lambda_{(n)}-\lambda_{(m)}$ and an
inequality for its norm in $W_{3}^{p}$ for each $t\in\lbrack t_{0},T].$ We
thus have shown the convergence of the sequence to a limit $\lambda$ which
satisfies the required equation. One can prove similarly that $\lambda\in
E_{3}^{p}(T).$ The uniqueness of the solution results from the monotony of $%
f(\lambda).$

\paragraph{Bounds for $\protect\lambda$}

When $\lambda\in C^{2}$ one obtains a lower bound by using the maximum
principle: at a minimum of $\lambda$ we have $\Delta\lambda\geq0.$ Hence a
minimum $\lambda_{m}$ of $\lambda$ satisfies the inequality 
\begin{equation*}
e^{2\lambda_{m}}\geq\frac{2}{\tau^{2}},\text{ \ \ \ i.e. \ \ \ }%
e^{-2\lambda_{m}}\leq\frac{1}{2}\tau^{2}
\end{equation*}
when $\lambda$ $\in E_{3}^{p}$ is solution of the equation it satisfies the
same inequality since $W_{3}^{p}\subset C^{0}$ and $\lambda$ can be obtained
as a limit in $W_{3}^{p}$ of functions satisfying this inequality.

An analogous argument shows that 
\begin{equation*}
\lambda_{-}\leq\lambda\leq\lambda_{+}
\end{equation*}
with 
\begin{equation*}
\lambda_{-}=\omega-maxv+v,\text{ and }\lambda_{+}=\omega+v-minv
\end{equation*}
where $v\in E_{2}\cap E_{3}^{p}$ is the solution with mean value zero on $%
\Sigma_{t}$ of the linear equation 
\begin{equation*}
\Delta v=f(\omega)\equiv p_{1}e^{2\omega}-p_{2}e^{-2\omega}+p_{3}
\end{equation*}
with $e^{2\omega}$ the positive solution of the equation

\begin{equation*}
P_{1}e^{4\omega}+P_{3}e^{2\omega}-P_{2}=0
\end{equation*}

\paragraph{Case of initial values.}

The above construction applies in particular on the initial surface $%
\Sigma_{0}.$ In this case the functions $u_{0}$ and $\overset{.}{u}_{0}$ are
considered as given. We have 
\begin{equation*}
\Delta v_{0}=f(\omega_{0})\equiv p_{1,0}e^{2\omega_{0}}-p_{2,0}e^{-2\omega
_{0}}+p_{3,0}
\end{equation*}
with 
\begin{equation*}
p_{1,0}=\frac{1}{4}\tau_{0}^{2},p_{2,0}=\frac{1}{2}(\mid h_{0}\mid^{2}+\mid%
\dot{u}_{0}\mid^{2}),p_{3,0}=-\frac{1}{2}(1+\mid Du_{0}\mid^{2})
\end{equation*}
We have 
\begin{equation*}
e^{2\omega_{0}}=\frac{(V_{\sigma}+\parallel Du_{0}\parallel^{2})+\sqrt
{(V_{\sigma}+\parallel Du_{0}\parallel^{2})^{2}+2\tau_{0}^{2}(\parallel
h_{0}\parallel^{2}+\parallel\dot{u}_{0}\parallel^{2})}}{V_{\sigma}%
\tau_{0}^{2}}
\end{equation*}
we see that $e^{2\omega_{0}}$ tends to $\frac{2}{\tau_{0}^{2}}$ and $%
\parallel f(\omega_{0})\parallel$ tends to zero when $q_{0}$ tends to zero
as well as the $H_{1}$ norms of $Du_{0}$ and $\overset{.}{u}_{0}$ (then the $%
L^{2}$ norm of $h_{0}$ tends also to zero)$.$

\subsubsection{Determination of the lapse N.}

The lapse N satisfies the equation 
\begin{equation}
\Delta N-\alpha N=-e^{2\lambda}\partial_{t}\tau
\end{equation}
with 
\begin{equation*}
\alpha e^{-2\lambda}=\frac{1}{2}\tau^{2}+e^{-4\lambda}(\mid\dot{u}\mid
^{2}+\mid h\mid^{2})>0
\end{equation*}

It is a well posed elliptic equation on $(\Sigma_{t},\sigma_{t}),$ when $u,h$
and $\lambda$ are known,which has one and only one solution, always
positive, in $E_{4}(T)$ in case a, in $E_{3}^{p}(T)$ in case b. Indeed:

\subparagraph{Case a.}

We have $\overset{.}{u}\in E_{2}(T),h\in E_{3}(T),\lambda\in E_{4}(T)$ hence
also $e^{2\lambda}\in E_{4}(T)$ and $\alpha\in E_{2}(T).$ The equation has
then a solution $N\in E_{4}(T).$

\subparagraph{Case b.}

We have $|\overset{.}{u}|^{2}+|h|^{2}\in E_{1}^{p}(T),e^{2\lambda
},e^{-2\lambda}\in E_{3}^{p}(T).$ The equation has a solution $N\in
E_{3}^{p}(T).$

\paragraph{Upper bound of N.}

At a maximum $x_{M}$ of N $\in C^{2}$ we have ($\Delta N)(x_{M})\leq0$ hence
this maximum $N_{M}$ \ is such that 
\begin{equation*}
N_{M}\leq(\alpha^{-1}e^{2\lambda}\partial_{t}\tau)(x_{M})
\end{equation*}
a fortiori

\begin{equation*}
N_{M}\leq\frac{2\partial_{t}\tau}{\tau^{2}}
\end{equation*}
A reasoning analogous to that given for $\lambda$ shows that this upper
bound also holds in case b.

\subsubsection{Determination of the shift $\protect\nu.$}

The definition of $k$ implies that $n\equiv e^{-2\lambda}\nu$ satisfies a
linear differential equation involving an operator L, the conformal Lie
derivative, with injective symbol: 
\begin{equation*}
(L_{\sigma_{t}}n)_{ab}\equiv
D_{a}n_{b}+D_{b}n_{a}-\sigma_{ab}D_{c}n^{c}=f_{ab}
\end{equation*}
with 
\begin{equation*}
f_{ab}\equiv2Ne^{-2\lambda}h_{ab}+\partial_{t}\sigma_{ab}-\frac{1}{2}%
\sigma_{ab}\sigma^{cd}\partial_{t}\sigma_{cd}
\end{equation*}
The kernel of the dual of $L$ is the space of transverse traceless symmetric
2-tensors in the metric $\sigma_{t},$ the equation for $\nu$ admits a
solution if and only if $f$ is L$^{2}-$orthogonal to all such tensors, i.e. 
\begin{equation*}
\int_{\Sigma_{t}}f_{ab}X_{I}^{ab}\mu_{\sigma_{t}}=0,\text{ for I =1,...6G-6}
\end{equation*}
This integrability condition will not in general be satisfied with the
arbitrary choice of $P(t)$, that is of $q_{t}\equiv h_{t}^{TT}$. In this
subsection $P(t)$ is not considered as given.We set in the expression of $%
f_{ab\text{ }}$%
\begin{equation*}
h_{t}=P^{I}(t)X_{I}(Q(t))+r_{t}
\end{equation*}
When $\sigma_{t}$ is a known $C^{1}$ function of $t$ the intgrability
condition determines $P(t)$ as a continuous field of tangent vectors to $%
T_{eich}$ by an invertible system of ordinary linear equations.

When $h$ is so chosen the equation for $n$ has a solution, unique since $%
L_{\sigma}$ has a trivial kernel on manifolds with $R(\sigma)=-1.$ It
results from elliptic theory that $n\in E_{4}(T)$ in case a, and $n\in
E_{3}^{p}(T)$ in case b.The same properties hold for $\nu.$

\subsection{Wave equation, local solution.}

The wave equation on ($\Sigma\times R)$ in the metric $^{(3)}g$ reads 
\begin{equation*}
-N^{-1}\partial_{0}(N^{-1}\partial_{0}u)+Ng^{ab}\nabla_{a}(N\partial
_{b}u)+N^{-1}\tau\partial_{0}u=0
\end{equation*}
We suppose that $\sigma_{t}$ is a given regular riemannian metric for $%
t\in\lbrack t_{0},T]$ and that $\lambda,N,\nu$ are given in $E_{3}^{p}(T)$
with $p>1$ and $N>0.$ Then we have $^{(3)}g\in E_{3}^{p}(T)$ $\subset
C^{1}(\Sigma\times\lbrack t_{0},T])$ and $^{(3)}g$ has hyperbolic signature.
It is easy to prove along standard lines that the Cauchy problem with data $%
u_{0},(\partial_{t}u)_{0}\in H_{2}\times H_{1}$ has a solution such that ($%
u,\partial_{t}u)\in E_{2}(T)\times E_{1}(T)$ on $\Sigma\times\lbrack
t_{0},T].$ The initial value $(\partial_{t}u)_{0}$ is the product of the
datum $\overset{.}{u}_{0}$ by $e^{-2\lambda_{0}},$ it belongs to $H_{1}$
under the hypothesis made in section 1.1 on the Cauchy data.

\subsection{Teichmuller parameters.}

We suppose known $h\in E_{2}^{p}(T),\lambda,N,\nu\in E_{3}^{p}(T)$, $u\in
E_{2}(T),$ and we suppose given $Q\rightarrow\psi(Q)$ a smooth cross section
of $M_{-1}$ over $T_{eich}.$ The unknown is the curve $t\rightarrow Q(t).$
We have $\sigma_{t}\equiv\psi(Q(t))$ and 
\begin{equation*}
\partial_{t}\sigma_{ab}=\frac{dQ}{dt}^{I}X_{I,ab}+C_{ab}
\end{equation*}
with $X_{I}(Q)$ a basis of the space of TT tensors on ($\Sigma,\psi(Q))$ and 
$C$ a conformal Lie derivative, $L^{2}$ orthogonal to TT tensors. The curve $%
t\rightarrow Q(t)$ and the tangent vector $P^{I}(t)$ to $T_{eich}$ satisfy
the ordinary differential system (cf. section 2.3.3) 
\begin{equation*}
X_{IJ}\frac{dQ^{I}}{dt}+Y_{IJ}P^{I}+Z_{J}=0
\end{equation*}
and 
\begin{equation*}
X_{IJ}\frac{dP^{I}}{dt}+\Phi_{J}(P,\frac{dQ}{dt})=0
\end{equation*}

This quasi linear first order system for $P$ and $Q$ has coefficients
continuous in $t$ and smooth in $Q$ and $P.$ The matrix of the principal
terms, $X_{IJ},$ is invertible. There exists therefore a number $T>0$ such
that the system has one and only one solution in $C^{1}([t_{0},T])$ with
given initial data $P_{0},Q_{0}.$

\subsection{Local existence theorem.}

We can now prove the following theorem

\begin{theorem}
The Cauchy problem with data ($u_{0,}\overset{.}{u}_{0})\in H_{2}\times
H_{1} $, on $\Sigma_{t_{0}}$ (denoted $\Sigma_{0})$ and $Q_{0},$ a point in $%
T_{eich},P_{0}$ a tangent vector to $T_{eich},$ for the Einstein equations
with U(1) isometry group (polarized case) has a solution with $\sigma_{t}$ a
regular metric on $\Sigma_{t}$ for t$\in\lbrack t_{0},T]$ and $u\in$ $%
E_{2}(T),T>t_{0},$ if $T-t_{0\text{ }}$is small enough. This solution is
unique when $\tau$ , depending only on $t,$ is chosen together with a cross
section of $M_{-1}$ over $T_{eich}.$

\begin{remark}
One has, for this solution, $\lambda,N,\nu\in E_{3}^{p}(T),1<p<2$ and $N>0. $
\end{remark}
\end{theorem}

Proof. The proof is straightforward, using iteration to solve alternatively
the elliptic systems, the wave equation and the differential system
satisfied by Teichmuller parameters, with $\tau$ a given function of $t$ and 
$\sigma _{t}$ required to remain in a chosen cross section of $M_{-1}$ over $%
T_{eich}.$ The iteration converges if $T-t_{0}$ is small enough. The limit
can be shown to be a solution of Einstein equations with $^{(3)}g$ in
constant mean curvature gauge by standard arguments, the 2-metric $g$ is
conformal with the factor $e^{2\lambda}$ to a metric in the chosen cross
section by construction.

This local existence theorem can be extended to the non polarized case.

\section{Scheme for a global existence theorem.}

As it is well known we will deduce from our local existence theorem a global
one, i.e. on $\Sigma\times\lbrack t_{0},\infty),$ if we can prove that the
curve $Q(t)$ remains in a compact subset of $T_{eich}$ and that neither the $%
H_{2}\times H_{1}$ norm of $(u(.,t),\overset{.}{u}(.,t))$ nor the $E_{3}^{p}$
norms of $\lambda(.,t),N(.,t),\nu(.,t)$ blow up when $t\in$ $[t_{0,}\infty)$
while $N$ remains strictly positive.

If the spacetime we construct is supported by the manifold $M\times\lbrack
t_{0},\infty)$ it will reach a moment of maximum expansion. It will be after
an infinite proper time for observers moving along orthogonal trajectories
of the hypersurfaces $M_{t}\equiv M\times\{t\}$ if the lapse function is
uniformly bounded below by a strictly positive number.

Our proof of this fact will rely on various refined estimates, using in
particular corrected energies. The correction of the energies poses special
problems in the non polarized case, which we will treat in another paper.

\subsection{Notations.}

$|.|$ and \TEXTsymbol{\vert}.\TEXTsymbol{\vert}$_{g}$ : pointwise norms of
scalars or tensors on $\Sigma,$ in the $\sigma$ or $g$ metric

$\parallel.\parallel$ and $\parallel.\parallel_{p}$: $L^{2}$ and $L^{p}$
norms in the $\sigma$ metric

$\parallel.\parallel_{g}$: $L^{2}$ norm in the $g$ metric.

A lower case index m or M denote respectively the lower or upper bound of a
scalar function on $\Sigma_{t}$. It may depends on t.

When we have to make a choice of the time parameter $t$ we will set 
\begin{equation}
t=-\tau^{-1}
\end{equation}
then $t$ will increase from $t_{0}>0$ to infinity when, $\Sigma_{t}$
expanding, $\tau(t)$ increases from $\tau_{0}<0$ to zero. With this choice
the upper bound on $N$ of subsection 4.3.4 reads 
\begin{equation}
N\leq2.
\end{equation}

\begin{remark}
Other admissible choices of $t$, for instance $\tau=t,$ $t\in\lbrack
t_{0},0),$ $t_{0}=\tau_{0}<0,$ would lead to the same geometrical
conclusions.
\end{remark}

\subsection{ Fundamental inequalities.}

\textbf{Lemma 1. }Let f be a scalar function on $\Sigma$. the following
inequalities hold

1.

\begin{equation*}
\parallel f\parallel_{q}\leq e^{-2\lambda_{m}/q}\parallel f\parallel
_{L^{q}(g)}
\end{equation*}
\begin{equation*}
\parallel f\parallel_{L^{q}(g)}\leq e^{2\lambda_{M}/q}\parallel
f\parallel_{q}
\end{equation*}

2. 
\begin{equation*}
|Df|_{g}=e^{-\lambda}|Df|,\text{ \ \ }\parallel
Df\parallel_{L^{\infty}(g)}\leq e^{-\lambda_{m}}\parallel
Df\parallel_{\infty}
\end{equation*}
and if $\ q\geq2$%
\begin{equation*}
\parallel Df\parallel_{L^{q}(g)}\leq e^{-\lambda_{m}(q-2)/q}\parallel
Df\parallel_{q}
\end{equation*}
in particular\ 
\begin{equation*}
\text{\ \ }\parallel Df\parallel=\parallel Df\parallel_{g}
\end{equation*}

3a. 
\begin{equation*}
\text{\ }|D^{2}f|=e^{2\lambda}|D^{2}f|_{g}
\end{equation*}
\begin{equation*}
\parallel D^{2}f\parallel_{L^{q}(g)}\leq e^{-2\frac{q-1}{q}%
\lambda_{m}}\parallel D^{2}f\parallel_{q}
\end{equation*}

3b.. 
\begin{equation*}
\Vert D^{2}f\Vert\leq e^{\lambda_{M}}\Vert\Delta_{g}f\Vert_{g}+\frac{1}{%
\sqrt{2}}\Vert Df\Vert_{g}
\end{equation*}

Proof: The inequalities 1, 2,3.a are trivial consequences of the identities: 
\begin{equation*}
\int_{\Sigma}f^{q}\mu_{\sigma}=\int_{\Sigma}f^{q}e^{-2\lambda}\mu_{g}\text{
\ since \ }\mu_{g}=e^{2\lambda}\mu_{\sigma}
\end{equation*}
and 
\begin{equation*}
g^{ab}D_{a}fD_{b}f=e^{-2\lambda}\sigma^{ab}D_{a}fD_{b}f
\end{equation*}
and a corresponding equality for $D^{2}f$ or, more generally, for covariant
2-tensors

To prove 3b we use the identity obtained by two successive partial
integrations and the Ricci formula with $R(\sigma)=-1$%
\begin{equation*}
\Vert D^{2}u\Vert^{2}=\Vert\Delta u\Vert^{2}+\frac{1}{2}\Vert Du\Vert^{2}
\end{equation*}
We have 
\begin{equation*}
\Delta u=e^{2\lambda}\Delta_{g}u,\text{ and }\parallel e^{2\lambda}\Delta
_{g}u\parallel=\parallel e^{\lambda}\Delta_{g}u\parallel_{g}
\end{equation*}
The given result follows.

\textbf{Lemma 2.}

We denote by $C_{\sigma}$ any positive number depending only on $(\Sigma
,\sigma).$

1.\ \textbf{\ }Let f be a scalar function on $\Sigma.$ There exists $%
C_{\sigma}$ such that the $L^{4}$ norms of $f$ and $Df$ are estimated by: 
\begin{equation*}
\Vert f\Vert_{4}\leq C_{\sigma}\{e^{-\lambda_{m}}\Vert f\Vert_{g}+e^{-\frac
{1}{2}\lambda_{m}}\Vert f\Vert_{g}^{\frac{1}{2}}\Vert Df\Vert_{g}^{\frac{1}{2%
}})
\end{equation*}
and 
\begin{equation*}
\Vert Df\Vert_{4}\leq C_{\sigma}\{\Vert Df\Vert_{g}+\Vert Df\Vert_{g}^{\frac{%
1}{2}}e^{\frac{1}{2}\lambda_{M}}\Vert\Delta_{g}f\Vert_{g}^{\frac{1}{2}})
\end{equation*}

2. For any $q$ such that $1\leq q<\infty$ there exists $C_{\sigma}$ such
that 
\begin{equation*}
\parallel f\parallel_{q}\leq C_{\sigma}\parallel f\parallel_{H_{1}}
\end{equation*}

Proof. 1. By the Sobolev inequalities there exists $C_{\sigma}$ such that 
\begin{equation*}
\parallel f\parallel_{4}^{2}=\Vert\mid f\mid^{2}\Vert\leq
C_{\sigma}(\Vert\mid f\mid^{2}\Vert_{1}+\Vert D\mid f\mid^{2}\Vert_{1})
\end{equation*}
Using 
\begin{equation*}
D\mid f\mid^{2}=2f.Df
\end{equation*}
we obtain 
\begin{equation*}
\Vert\mid f\mid^{2}\Vert\leq C_{\sigma}\Vert f\Vert(\Vert f\Vert+2\Vert
Df\Vert)
\end{equation*}
which gives the first result using the lemma 1.

Analogously 
\begin{equation*}
\parallel Df\parallel_{4}^{2}\equiv\parallel|Df|^{2}\parallel\leq C_{\sigma
}\{\parallel Df\parallel^{2}+\parallel D|Df|^{2}\parallel_{1}\}
\end{equation*}
leads to the second inequality.

2.\ The Sobolev embedding theorem and the compactness of $\Sigma.$

\section{Energy estimates.}

\subsection{Bound of the first energy.}

The 2+1 dimensional Einstein equations with source the stress energy tensor
of the wave map u contain the following equation (hamiltonian constraint) 
\begin{equation}
2N^{-2}(T_{00}-^{(3)}S_{00})=N^{-2}\partial_{0}u.\partial_{0}u+g^{ab}%
\partial_{a}u.\partial_{b}u+g^{ab}g^{cd}k_{cb}k_{da}-R-\tau^{2}=0
\end{equation}

Recall the splitting of the covariant 2-tensor $k$ into a trace and a
traceless part: 
\begin{equation}
k_{ab}=h_{ab}+\frac{1}{2}g_{ab}\tau
\end{equation}
hence

\begin{equation}
|k|_{g}^{2}=g^{ac}g^{bd}k_{ab}k_{cd}=|h|_{g}^{2}+\frac{1}{2}\tau^{2}
\end{equation}
and the hamiltonian constraint equation reads 
\begin{equation}
|u^{\prime}|^{2}+|Du|_{g}^{2}+|h|_{g}^{2}=R(g)+\frac{1}{2}\tau^{2}
\end{equation}
with 
\begin{equation*}
u^{\prime}\equiv N^{-1}\partial_{0}u
\end{equation*}

We define the first energy by the following formula (recall that $|.|_{g}$
and $\parallel.\parallel_{g}$denote respectively the pointwise norm and the $%
L^{2}$ norm in the metric $g)$

\begin{equation}
E(t)=\frac{1}{2}\int_{\Sigma_{t}}(|u^{\prime}|^{2}+|Du|_{g}^{2}+|h|_{g}^{2})%
\mu_{g}\equiv\frac{1}{2}\{\parallel u^{\prime}\parallel_{g}^{2}+\parallel
Du\parallel_{g}^{2}+\parallel h\parallel_{g}^{2}\}
\end{equation}
This energy is the first energy of the wave map $u$ completed by the $%
L^{2}(g)$ norm of $h.$

We integrate the hamiltonian constraint on ($\Sigma_{t,}g)$ using the
constancy of $\tau$ and the Gauss Bonnet theorem which reads, with $\chi$
the Euler characteristic of $\Sigma$

\begin{equation*}
\int_{\Sigma_{t}}R(g)\mu_{g}=4\pi\chi
\end{equation*}
We have then 
\begin{equation}
E(t)=\frac{\tau^{2}}{4}Vol_{g}(\Sigma_{t})+2\pi\chi
\end{equation}

\noindent with

\begin{equation*}
Vol_{g}(\Sigma_{t})=\int_{\Sigma_{t}}\mu_{g}
\end{equation*}

We know from elementary calculus that on a compact manifold 
\begin{equation*}
\frac{dVol_{g}\Sigma_{t}}{dt}=\frac{1}{2}\int_{\Sigma_{t}}g^{ab}\frac{%
\partial g_{ab}}{\partial t}\mu_{g}=-\tau\int_{\Sigma_{t}}N\mu_{g}
\end{equation*}
since 
\begin{equation*}
g^{ab}\partial_{t}g_{ab}=-2N\tau+2\nabla^{a}\nu_{a}
\end{equation*}
We use the equation 
\begin{equation*}
N^{-1(3)}R_{00}=\Delta_{g}N-N|k|_{g}^{2}+\partial_{t}\tau=|u^{\prime}|^{2}
\end{equation*}
together with the splitting of $k$ to write after integration, since $\tau$
is constant in space, 
\begin{equation*}
\frac{1}{2}\tau^{2}\int_{\Sigma_{t}}N\mu_{g}=\frac{d\tau}{dt}%
Vol_{g}(\Sigma_{t})-\int_{\Sigma_{t}}N(|h|_{g}^{2}+|u^{\prime}|^{2})\mu_{g}
\end{equation*}
We use these results to compute the derivative of $\ E(t)$ and we find that
it simplifies to: 
\begin{equation*}
\frac{dE(t)}{dt}=\frac{1}{2}\tau\int_{t}(|h|_{g}^{2}+|u^{\prime}|^{2})N\mu
_{g}.
\end{equation*}
We see that $E(t)$ is a non increasing function of t if $\tau$ is negative.
The absence of the term $|Du|_{g}^{2}$ \ on the right hand side does not
permit an estimate of the rate of decay of E(t).

We will estimate this decay in a forthcoming section.

Note in addition the appearance of $N$ in the right hand side.

\subsection{Second energy estimates}

In this paragraph indices are raised with $g.$ We denote by $h_{g}^{ab}$ the
contravariant components of $h_{ab}$ computed with the metric $g.$

We define the energy of gradient $u$ by the formula 
\begin{equation*}
E^{(1)}(t)\equiv\int_{\Sigma_{t}}(J_{0}+J_{1})\mu_{g}
\end{equation*}
with 
\begin{equation*}
J_{1}=\frac{1}{2}\mid\Delta_{g}u\mid^{2},\text{ \ \ }J_{0}=\frac{1}{2}\mid
Du^{\prime}\mid^{2}
\end{equation*}
We have for an arbitrary function $f:$%
\begin{equation*}
\frac{d}{dt}\int_{\Sigma_{t}}f\mu_{g}=\int_{\Sigma_{t}}\{\partial_{t}f+\frac{%
1}{2}g^{ab}\partial_{t}g_{ab}\}\mu_{g}
\end{equation*}
that is, due to the definition of $k_{ab},$%
\begin{equation*}
\frac{d}{dt}\int_{\Sigma_{t}}f\mu_{g}=\int_{\Sigma_{t}}\{\partial_{t}f-(N%
\tau-\nabla_{a}\nu^{a})f\}\mu_{g}
\end{equation*}
hence after integration by parts on the compact manifold $\Sigma,$ using the
expression of $\partial_{0}$ and replacing $f$ by $J_{0}+J_{1}$ the
following formula where the shift does not appear explicitly: 
\begin{equation*}
\frac{d}{dt}\int_{\Sigma_{t}}(J_{1}+J_{0})\mu_{g}=\int_{\Sigma_{t}}\{%
\partial_{0}(J_{1}+J_{0})-N\tau(J_{1}+J_{0})\}\mu_{g}
\end{equation*}

We first compute 
\begin{equation*}
\int_{\Sigma_{t}}\partial_{0}J_{1.}\mu_{g}=\int_{\Sigma_{t}}\partial_{0}%
\Delta_{g}u.\Delta_{g}u\mu_{g}
\end{equation*}
We define the operator ${\bar{\partial}}_{0}$ on time dependent space
tensors by 
\begin{equation*}
{\bar{\partial}}_{0}=\partial_{0}-L_{\nu}
\end{equation*}
where $L_{\nu}$ denotes the Lie derivative in the direction of the shift $%
\nu.$We have 
\begin{equation*}
{\overline{\partial}}_{0}\Delta_{g}u=g^{ab}{\overline{\partial}}_{0}\nabla
_{a}\partial_{b}u+{\overline{\partial}}_{0}g^{ab}\nabla_{a}\partial_{b}u
\end{equation*}
Therefore using 
\begin{equation*}
\overset{\_}{\partial_{0}}g^{ab}=2Nk^{ab}\equiv2Nh_{g}^{ab}+Ng^{ab}\tau
\end{equation*}
\begin{equation*}
\int_{\Sigma_{t}}\partial_{0}J_{1}\mu_{g}=\int_{\Sigma_{t}}g^{ab}{\overline{%
\partial}}_{0}\nabla_{a}\partial_{b}u.\Delta_{g}u\mu_{g}+X_{1}
\end{equation*}
with 
\begin{equation*}
X_{1}=\int_{\Sigma_{t}}\{2Nh_{g}^{ab}\nabla_{a}\partial_{b}u.\Delta
_{g}u+2N\tau J_{1}\}\mu_{g}
\end{equation*}

Analogously 
\begin{equation*}
\int_{\Sigma_{t}}\partial_{0}J_{0}\mu_{g}=\int_{\Sigma_{t}}g^{ab}\overset{\_%
}{\partial_{0}}\partial_{a}u^{\prime}.\partial_{b}u^{\prime}\mu_{g}+X_{0}
\end{equation*}
with 
\begin{equation*}
X_{0}=\int_{\Sigma_{t}}\{Nh_{g}^{ab}\partial_{a}u^{\prime}.\partial
_{b}u^{\prime}+N\tau J_{0}\}\mu_{g}
\end{equation*}

We use the commutation of the operator $\overset{\_}{\partial}_{0}$ with the
partial derivative $\partial_{a}$ (cf. C.B-York 1995) together with partial
integration to obtain

\begin{center}
$\int_{\Sigma_{t}}g^{ab}\overset{\_}{\partial_{0}}\partial_{a}u^{\prime
}.\partial_{b}u^{\prime}\mu_{g}=-\int_{\Sigma_{t}}\partial_{0}u^{\prime
}.\Delta_{g}u^{\prime}\mu_{g}$
\end{center}

The function $u$ satisfies the wave equation on $(\Sigma\times R,^{(3)}g),$
namely: 
\begin{equation*}
\partial_{0}u^{\prime}=N\Delta_{g}u+\partial^{a}N\partial_{a}u+\tau
Nu^{\prime}
\end{equation*}
which gives 
\begin{equation*}
\int_{\Sigma_{t}}g^{ab}\overset{\_}{\partial_{0}}\partial_{a}u^{\prime
}.\partial_{b}u^{\prime}\mu_{g}=-\int_{\Sigma_{t}}N\Delta_{g}u.\Delta
_{g}u^{\prime}\mu_{g}+Y_{0}
\end{equation*}
with, after another integration by parts 
\begin{equation*}
Y_{0}\equiv\int_{\Sigma_{t}}\{(\nabla_{b}(\partial^{a}N\partial_{a}u)+\tau%
\partial_{b}Nu^{\prime}).(\partial^{b}u^{\prime})+2\tau NJ_{0}\}\mu_{g}
\end{equation*}

On the other hand: 
\begin{equation*}
g^{ab}\overset{\_}{\partial}_{0}\nabla_{a}\partial_{b}u\equiv\Delta
_{g}\partial_{0}u-g^{ab}\overset{\_}{\partial}_{0}\Gamma_{ab}^{c}%
\partial_{c}u
\end{equation*}
with 
\begin{equation*}
\Delta_{g}\partial_{0}u\equiv\Delta_{g}(Nu^{\prime})\equiv N\Delta u^{\prime
}+2\partial^{a}\partial_{a}u^{\prime}+u^{\prime}\Delta_{g}N
\end{equation*}
therefore 
\begin{equation*}
\int_{\Sigma_{t}}g^{ab}\overset{\_}{\partial}_{0}\nabla_{a}\partial
_{b}u.\Delta_{g}u\mu_{g}=\int_{\Sigma_{t}}N\Delta_{g}u.\Delta_{g}u^{\prime}%
\mu_{g}+Y_{1}
\end{equation*}
with 
\begin{equation*}
Y_{1}=\int_{\Sigma_{t}}\{-g^{ab}\overset{\_}{\partial}_{0}\Gamma_{ab}^{c}%
\partial_{c}u+2\partial^{a}N\partial_{a}u^{\prime}+u^{\prime}\Delta
_{g}N\}.\Delta_{g}u\mu_{g}
\end{equation*}
which can be written, using the identity

\begin{equation*}
{\overline{\partial}}_{0}\Gamma_{ab}^{c}=\nabla^{c}(Nk_{ab})-%
\nabla_{a}(Nk_{b}^{c})-\nabla_{b}(Nk_{a}^{c})
\end{equation*}
together with the equation 
\begin{equation*}
\nabla_{a}k_{b}^{a}=-\partial_{b}u.u^{\prime}
\end{equation*}
\begin{equation}
Y_{1}=\int_{\Sigma_{t}}\{(2\partial_{a}Nh_{g}^{ac}-2N\partial^{c}u.u^{\prime
})\partial_{c}u+2\partial^{a}N\partial_{a}u^{\prime}+u^{\prime}\Delta
_{g}N\}.\Delta_{g}u\mu_{g}
\end{equation}
We see that the terms in third derivatives of $u$ disappear in the
derivative of $E^{(1)}(t).$ We have obtained 
\begin{equation*}
\int_{\Sigma_{t}}\partial_{0}(J_{0}+J_{1})\mu_{g}=X_{0}+X_{1}+Y_{0}+Y_{1}
\end{equation*}
where the X's and Y' are given by the above formulas. We read from these
formulas the following theorem

\begin{theorem}
The time derivative of the second energy $E^{(1)}$ satisfies the equality
\end{theorem}

\begin{equation}
\frac{dE^{(1)}}{dt}-2\tau
E^{(1)}=\tau\int_{\Sigma_{t}}NJ_{0}+(N-2)(J_{0}+J_{1})\mu_{g}++Z
\end{equation}
The quantity $Z$ is given by: 
\begin{align}
Z & \equiv\int_{\Sigma_{t}}\{Nh_{g}^{ab}\partial_{a}u^{\prime}.\partial
_{b}u^{\prime}+2Nh_{g}^{ab}\nabla_{a}\partial_{b}u.\Delta_{g}u+ \\
& (\nabla_{b}(\partial^{a}N\partial_{a}u)+\tau\partial_{b}Nu^{\prime
}).(\partial^{b}u^{\prime})\}\mu_{g}+Y_{1}
\end{align}
For $\tau\leq0,$ and $0<N\leq2,$ the right hand side of (19) is less than $%
Z, $ which can be estimated with non linear terms in the energies: all the
terms which are only quadratic in the derivatives of u, i.e. linear in
energy densities, have coefficients which contain $N-2,$ $\partial_{a}N$ or $%
h_{g}^{ab}$, or their derivatives. To estimate these terms we need bounds
which will be deduced from estimates on the conformal factor and the lapse $%
N $.

In the following paragraphs we will set 
\begin{equation*}
E(t)\equiv\varepsilon^{2},\text{ \ \ and \ \ }\tau^{-2}E^{(1)}(t)\equiv
\varepsilon_{1}^{2}
\end{equation*}

\section{Estimates for h in $H_{1}.$}

\subsection{Estimate of $\parallel h\parallel.$\protect\bigskip}

We have defined the auxiliary unknown $h$ by 
\begin{equation*}
h_{ab}\equiv k_{ab}-\frac{1}{2}g_{ab}\tau
\end{equation*}
Its $L^{2}$ norm on $(\Sigma,\sigma)$ is bounded in terms of the first
energy and an upper bound $\lambda_{M}$ of the conformal factor since we
have 
\begin{equation*}
\parallel
h\parallel^{2}=\int_{\Sigma_{t}}\sigma^{ac}\sigma^{bd}h_{ab}h_{cd}\mu_{%
\sigma}=\int_{\Sigma_{t}}e^{2\lambda}g^{ac}g^{bd}h_{ab}h_{cd}\mu_{g}\leq
e^{2\lambda_{M}}\parallel h\parallel_{L^{2}(g)}^{2}
\end{equation*}
which implies on $\Sigma_{t},$ by the definition of $E(t),$ 
\begin{equation*}
\parallel h\parallel\leq e^{\lambda_{M}}\varepsilon
\end{equation*}
with 
\begin{equation*}
\varepsilon\equiv E^{\frac{1}{2}}(t)
\end{equation*}

\subsection{Estimate of $\parallel Dh\parallel.$\protect\bigskip}

The tensor $h$ satisfies the equations 
\begin{equation*}
D_{a}h_{b}^{a}=L_{b}\equiv-\partial_{a}u.\overset{.}{u}
\end{equation*}
It is the sum of a TT tensor $h_{TT}\equiv q$ and a conformal Lie derivative 
$r$: 
\begin{equation*}
h\equiv q+r
\end{equation*}
It results from elliptic theory that on each $\Sigma_{t}$ the tensor r
satisfies the estimate 
\begin{equation*}
\parallel r\parallel_{H_{1}}\leq C_{\sigma}\parallel Du.\overset{.}{u}%
\parallel\leq C_{\sigma}\parallel|Du|^{2}\parallel^{\frac{1}{2}}\parallel|%
\overset{.}{u}|^{2}\parallel^{\frac{1}{2}}
\end{equation*}
We will bound the right hand side of this inequality in terms of the first
and second energies of $u$ . We have: 
\begin{equation*}
\parallel|\overset{.}{u}|^{2}\parallel\leq e^{4\lambda_{M}}\parallel
|u^{\prime}|^{2}\parallel
\end{equation*}
we have proven in section 4 that 
\begin{equation*}
\Vert\mid u^{\prime}\mid^{2}\Vert\leq C_{\sigma}e^{-\lambda_{m}}\Vert
u^{\prime}\Vert_{L^{2}(g)}(e^{-\lambda_{m}}\Vert
u^{\prime}\Vert_{L^{2}(g)}+\Vert Du^{\prime}\Vert_{L^{2}(g)})
\end{equation*}
We have set 
\begin{equation*}
\varepsilon_{1}\equiv|\tau|^{-1}\{E^{(1)}(t)\}^{\frac{1}{2}}
\end{equation*}
hence, using the lower bound on $\lambda$ and the definitions of $%
\varepsilon $ and $\varepsilon_{1}$ we obtain 
\begin{equation*}
\Vert\mid u^{\prime}\mid^{2}\Vert\leq C_{\sigma}\tau^{2}(\varepsilon
^{2}+\varepsilon\varepsilon_{1})
\end{equation*}
On the other hand 
\begin{equation*}
\Vert\mid Du\mid^{2}\Vert\leq C_{\sigma}\Vert Du\Vert_{L^{2}(g)}(\Vert
Du\Vert_{L^{2}(g)}+e^{\lambda_{M}}\Vert\Delta_{g}u\Vert_{L^{2}(g)})
\end{equation*}
hence 
\begin{equation*}
\Vert\mid Du\mid^{2}\Vert\leq C_{\sigma}\{\varepsilon^{2}+\varepsilon
\varepsilon_{1}e^{\lambda_{M}}|\tau|\}
\end{equation*}
It results from these inequalities that 
\begin{equation*}
\parallel r\parallel_{H_{1}}^{2}\leq C_{\sigma}e^{4\lambda_{M}}\tau
^{2}\varepsilon^{2}(\varepsilon+\varepsilon_{1}\}\{\varepsilon+\varepsilon
_{1}e^{\lambda_{M}}|\tau|\}
\end{equation*}

We now estimate the transverse part $h_{TT}=q.$

It is known (cf. Andersson and Moncrief ) that in dimension 2 the equation 
\begin{equation*}
D_{a}q_{b}^{a}=0,\text{ with }q_{a}^{a}=0
\end{equation*}
implies 
\begin{equation*}
D^{c}D_{c}q_{ab}=R(\sigma)q_{ab}.
\end{equation*}
When R($\sigma)=-1$ this equation gives by integration on $\Sigma_{t\text{ }%
} $ of its contracted product with $q^{ab}$ the following relation 
\begin{equation*}
\parallel Dq\parallel=\parallel q\parallel
\end{equation*}
more generally any $H_{s}$ norm of $q$ is a multiple of its $L^{2}$ norm.

We have 
\begin{equation*}
\parallel q\parallel\leq\parallel h\parallel+\parallel r\parallel
\end{equation*}
therefore 
\begin{equation*}
\parallel Dh\parallel\leq\parallel Dq\parallel+\parallel Dr\parallel
\leq\parallel h\parallel+\parallel r\parallel_{H_{1}}
\end{equation*}
In other words 
\begin{equation*}
\parallel Dh\parallel\leq
e^{\lambda_{M}}\varepsilon\{1+C_{\sigma}e^{\lambda_{M}}|\tau|(\varepsilon+%
\varepsilon_{1}e^{\lambda_{M}}|\tau |)^{\frac{1}{2}}(\varepsilon+%
\varepsilon_{1})^{\frac{1}{2}}\}
\end{equation*}

\section{Estimates for the conformal factor.}

\subsection{First estimates.}

Recall that we denote respectively by $\parallel.\parallel$ and $\Vert
.\Vert_{p}$ the $L^{2}(\sigma)$ and $L^{p}(\sigma)$ norms on $\Sigma$ and by 
$\parallel.\parallel_{g}$ an $L^{2}(g)$ norm on $\Sigma.$

The conformal factor $\lambda$ satisfies the equation 
\begin{equation*}
\Delta\lambda=f(\lambda)\equiv p_{1}e^{2\lambda}-p_{2}e^{-2\lambda}+p_{3}
\end{equation*}
where the coefficients $p_{i}$ are functions in $E_{0}\cap E_{1}^{p},1<p<2,$
hypothesis consistent with $Du,\overset{.}{u},h\in E_{1,}$ given by 
\begin{equation*}
p_{1}=\frac{1}{4}\tau^{2},p_{2}=\frac{1}{2}(\mid h\mid^{2}+\mid\dot{u}\mid
^{2}),p_{3}=\frac{1}{2}(R(\sigma)-\mid Du\mid^{2})
\end{equation*}
Having chosen $R(\sigma)=-1$ we have seen that a lower bound $\lambda_{m}$\
for $\lambda$ is such that 
\begin{equation*}
e^{-2\lambda_{m}}\leq\frac{1}{2}\tau^{2}
\end{equation*}

Also 
\begin{equation*}
\lambda_{-}\leq\lambda\leq\lambda_{+}
\end{equation*}
\begin{equation*}
\lambda_{-}=\omega-maxv+v,\text{ and }\lambda_{+}=\omega+v-minv
\end{equation*}
where $v\in E_{2}\cap E_{3}^{p}$ is the solution with mean value zero on $%
\Sigma_{t}$ of the linear equation 
\begin{equation*}
\Delta v=f(\omega)\equiv p_{1}e^{2\omega}-p_{2}e^{-2\omega}+p_{3}
\end{equation*}
where $e^{2\omega},$ positive solution of the equation

\begin{equation*}
P_{1}e^{4\omega}+P_{3}e^{2\omega}-P_{2}=0
\end{equation*}
is given by, since $P_{3}<0$, $P_{2}\geq0,$ \ $P_{1}=\frac{1}{4}\tau
^{2}V_{\sigma},$ 
\begin{equation*}
e^{2\omega}=\frac{-P_{3}+\sqrt{P_{3}^{2}+4P_{1}P_{2}}}{2P_{1}}\equiv \frac{%
-P_{3}(1+\sqrt{1+4P_{3}^{-2}P_{1}P_{2}})}{2P_{1}}
\end{equation*}
This formula will permit an estimate of $e^{2\omega}-\frac{2}{\tau^{2}},$ a
positive quantity, in terms of the energies. Indeed using the elementary
algebra inequality 
\begin{equation*}
\sqrt{1+a}\leq1+\frac{1}{2}a,\text{ when }a\geq0
\end{equation*}
we obtain 
\begin{equation*}
e^{2\omega}\leq-\frac{P_{3}}{P_{1}}-\frac{P_{2}}{P_{3}}
\end{equation*}
and, using the expressions of $P_{2},P_{3}$ and $P_{1}=\frac{1}{4}\tau
^{2}V_{\sigma},$ together with 
\begin{equation*}
\parallel\overset{.}{u}\parallel^{2}\leq e^{2\lambda_{M}}\parallel u^{\prime
}\parallel_{g}^{2},\text{ and }\parallel h\parallel^{2}\leq
e^{2\lambda_{M}}\parallel h\parallel_{g}^{2}
\end{equation*}
we find 
\begin{equation*}
0\leq\frac{1}{2}\tau^{2}e^{2\omega}-1\leq V_{\sigma}^{-1}\{\parallel
Du\parallel^{2}+\frac{\tau^{2}}{2}e^{2\lambda_{M}}(\parallel u^{\prime
}\parallel_{g}^{2}+\parallel h\parallel_{g}^{2})\}
\end{equation*}
We have set $\varepsilon^{2}\equiv E(t)$ and therefore we have

\begin{equation*}
0\leq\frac{1}{2}\tau^{2}e^{2\omega}-1\equiv\varepsilon_{\omega}\leq
V_{\sigma }^{-1}\{1+\frac{\tau^{2}}{2}e^{2\lambda_{M}})\varepsilon^{2}\}
\end{equation*}

We will now give estimates for $\lambda.$

\begin{lemma}
Denote by $\lambda_{M}$ the maximum of $\lambda,$ one has
\end{lemma}

\begin{equation*}
0\leq\lambda_{M}-\omega\leq2\parallel v\parallel_{L^{\infty}}
\end{equation*}
\begin{equation*}
0\leq\omega-\lambda_{m}\leq2\parallel v\parallel_{L^{\infty}}
\end{equation*}
Proof. The result follows from the expressions of $\lambda_{-}$ and $%
\lambda_{+}$ : 
\begin{equation*}
\lambda_{M}\leq\sup\lambda_{+}=\omega+maxv-minv,\text{ and }%
\lambda_{m}\geq\inf\lambda_{-}=\omega+minv-maxv
\end{equation*}
Also 
\begin{equation*}
\lambda_{M}-\lambda_{m}\leq2maxv-2minv\leq4maxv\leq4\parallel v\parallel
_{L^{\infty}}
\end{equation*}

\begin{corollary}
The following inequality holds
\end{corollary}

\begin{equation}
1\leq e^{\lambda_{M}-\omega}\leq1+2\parallel
v\parallel_{L^{\infty}}e^{2\parallel v\parallel_{L^{\infty}}}
\end{equation}

\begin{equation}
1\leq e^{\lambda_{M}-\lambda_{m}}\leq1+4\parallel
v\parallel_{L^{\infty}}e^{4\parallel v\parallel_{L^{\infty}}}
\end{equation}
Proof. Elementary calculus

We set 
\begin{equation*}
\varepsilon_{v}\equiv\parallel v\parallel_{L^{\infty}}
\end{equation*}
Denote by $\varepsilon_{v_{0}}$ the $L^{\infty}$ norm of the function $v$
computed with initial data. We have shown in the section on local existence
that $\varepsilon_{v_{0}}$ tends to zero with the initial data $q_{0},Du_{0}$
and $\overset{.}{u}_{0}.$

\textbf{Hypothesis H}$_{c}$. We say that v satisfies the hypothesis H$_{c}$
if there exists a number $c>\varepsilon_{v_{0}}$, independent of t, such
that $\varepsilon_{v}\leq c.$

We suppose also that the initial data are such that $E(t_{0})\equiv$ $%
\varepsilon_{0}^{2}$ verifies the inequality (we chose $\frac{1}{2}$ for
simplicity of notations)

\begin{equation*}
V_{\sigma_{0}}^{-1}\varepsilon_{0}^{2}(1+2ce^{2c})^{2}<\frac{1}{2}.
\end{equation*}
\textbf{\ }Then, since $E(t)$ is non increasing and the volume $V_{\sigma}$
of ($\Sigma,\sigma)$ is constant by the Gauss Bonnet theorem, it holds for
all $t$ that 
\begin{equation*}
V_{\sigma}^{-1}\varepsilon^{2}(1+2ce^{2c})^{2}<\frac{1}{2}.
\end{equation*}
\bigskip

Recall that we have denoted by $C_{\sigma}$ any positive number depending
only on $(\Sigma,\sigma).$

We denote by $C$ any positive number depending only on $c.$

\begin{theorem}
When $\varepsilon_{v}\leq c$ there exist numbers C such that the conformal
factor $\lambda$ satisfies the estimates:
\end{theorem}

1. 
\begin{equation*}
\frac{1}{2}\tau^{2}e^{2\lambda_{M}}\leq1+C(\varepsilon^{2}+\varepsilon_{v})
\end{equation*}

2. 
\begin{equation*}
e^{\lambda_{M}-\lambda_{m}}\leq1+C\varepsilon_{v}.
\end{equation*}

Proof.

1. We find, using the estimate of $\omega$: 
\begin{equation*}
1\leq\frac{1}{2}\tau^{2}e^{2\lambda_{M}}\equiv\frac{1}{2}e^{2(\lambda
_{M}-\omega)}\tau^{2}e^{2\omega}\leq(1+2\varepsilon_{v}e^{2%
\varepsilon_{v}})^{2}[1+V_{\sigma}{}^{-1}(1+\frac{\tau^{2}}{2}%
e^{2\lambda_{M}})\varepsilon ^{2}]
\end{equation*}
therefore 
\begin{equation*}
1\leq\frac{1}{2}\tau^{2}e^{2\lambda_{M}}\leq\frac{(1+2\varepsilon
_{v}e^{2\varepsilon_{v}})^{2}(1+V_{\sigma}^{-1}\varepsilon^{2})}{1-V_{\sigma
}^{-1}\varepsilon^{2}(1+2\varepsilon_{v}e^{2\varepsilon_{v}})^{2}}
\end{equation*}
that is 
\begin{equation*}
0\leq\frac{1}{2}\tau^{2}e^{2\lambda_{M}}-1\leq\frac{(1+2\varepsilon
_{v}e^{2\varepsilon_{v}})^{2}2V_{\sigma}^{-1}\varepsilon^{2}+4\varepsilon
_{v}e^{2\varepsilon_{v}}+4\varepsilon_{v}^{2}e^{4\varepsilon_{v}}}{%
1-V_{\sigma}^{-1}\varepsilon^{2}(1+2\varepsilon_{v}e^{2\varepsilon_{v}})^{2}}
\end{equation*}

The result 1 of the lemma follows then from the hypothesis H$_{c}$ and H$%
_{0}.$

2. Is immediate.

\subsection{Estimate of v.}

The equation satisfied by v implies 
\begin{equation*}
\int_{\Sigma}|Dv|^{2}\mu_{\sigma}=-\int_{\Sigma}f(\omega)v\mu_{\sigma}
\end{equation*}
hence 
\begin{equation*}
\parallel Dv\parallel^{2}\leq\parallel f(\omega)\parallel\parallel v\parallel
\end{equation*}
but the Poincare inequality applied to the function v which has mean value 0
on $\Sigma$ gives 
\begin{equation*}
\parallel v\parallel^{2}\leq\lbrack\Lambda]^{-1}\parallel Dv\parallel^{2}
\end{equation*}
where $\Lambda$ is the first (positive) eigenvalue of - $\Delta$ for
functions on $\Sigma_{t}$ with mean value zero. Therefore on each $%
\Sigma_{t} $%
\begin{equation*}
\parallel Dv\parallel\leq\lbrack\Lambda]^{-1/2}\parallel f_{0}\parallel
\end{equation*}
We use Ricci identity and $R(\sigma)=-1$ to obtain 
\begin{equation*}
\parallel\Delta v\parallel^{2}=\parallel D^{2}v\parallel^{2}-\frac{1}{2}%
\parallel Dv\parallel^{2}
\end{equation*}
The equation satisfied by v implies then 
\begin{equation*}
\Vert D^{2}v\Vert^{2}=\Vert f(\omega)\Vert^{2}+\frac{1}{2}\Vert Dv\Vert^{2}
\end{equation*}
Assembling these various inequalities implies 
\begin{equation*}
\parallel
v\parallel_{H_{2}}\leq\lbrack1+3/(2\Lambda)+1/\Lambda^{2}]^{1/2}\parallel
f(\omega)\parallel
\end{equation*}
The Sobolev inequality 
\begin{equation*}
\parallel v\parallel_{L^{\infty}}\leq C_{\sigma}\parallel v\parallel_{H_{2}}
\end{equation*}
gives then a bound on the $L^{\infty}$ norm of v on $\Sigma_{t\text{ }}$in
terms of the $L^{2}$ norm of $f(\omega)$, a Sobolev constant $C_{\sigma}$
and the lowest eigenvalue $\Lambda$ of $-\Delta$

We now estimate the $L^{2}$ norm of $f(\omega).$%
\begin{equation*}
f(\omega)\equiv f_{\omega}\equiv p_{1}e^{2\omega}-p_{2}e^{-2\omega}+p_{3}
\end{equation*}
We split $f_{\omega}$ into a constant part and a non constant part $%
h_{\omega }$ by setting 
\begin{equation*}
h_{\omega}\equiv p_{2}e^{-2\omega}+\frac{1}{2}|Du|^{2}.
\end{equation*}
Since the mean value $\overset{\_}{f}_{\omega}$ of $f(\omega)$ is zero and
the mean value of a constant is equal to itself we have 
\begin{equation*}
f_{\omega}\equiv\bar{h}_{\omega}-h_{\omega}.
\end{equation*}
By the isoperimetric inequality there exists a constant $I_{\sigma}$ such
that 
\begin{equation*}
\Vert f_{\omega}\Vert\leq I_{\sigma}\Vert Dh_{\omega}\Vert_{1}
\end{equation*}
We want to bound the right hand side in terms of the first and second
energies of the wave map. We have by the definition of $h_{\omega}:$%
\begin{equation*}
\Vert Dh_{\omega}\Vert_{1}\leq\frac{1}{2}\{\Vert
D|Du|^{2}\Vert_{1}+e^{-2\omega_{0}}(\Vert D|h|^{2}\Vert_{1}+\Vert D|\dot{u}%
|^{2}\Vert_{1})\}
\end{equation*}

\begin{lemma}
1. The following estimate holds 
\begin{equation*}
\frac{1}{2}\Vert D|Du|^{2}\Vert_{1}\leq\Vert
Du\Vert_{g}(e^{\lambda_{M}}\Vert\Delta_{g}u\Vert_{g}+(1/\sqrt{2})\Vert
Du\Vert_{g})
\end{equation*}
2.\ It implies under the hypothesis H$_{c}$ that 
\begin{equation*}
\frac{1}{2}\parallel D|Du|^{2}\parallel_{1}\leq
C(\varepsilon^{2}+\varepsilon\varepsilon_{1})
\end{equation*}
\end{lemma}

Proof. 1. We have: 
\begin{equation*}
D|Du|^{2}=2Du.D^{2}u
\end{equation*}
hence 
\begin{equation*}
\Vert D|Du|^{2}\Vert_{1}\leq2\Vert Du\Vert\Vert D^{2}u\Vert
\end{equation*}
Previous elementary calculus gave 
\begin{equation*}
\Vert Du\Vert=\parallel Du\parallel_{g}\equiv\Vert Du\Vert_{L^{2}(g)}
\end{equation*}
and 
\begin{equation*}
\Vert D^{2}u\Vert^{2}=\Vert\Delta u\Vert^{2}+\frac{1}{2}\Vert Du\Vert^{2}
\end{equation*}
with 
\begin{equation*}
\Delta u=e^{2\lambda}\Delta_{g}u,\text{ and }\parallel e^{2\lambda}\Delta
_{g}u\parallel=\parallel e^{\lambda}\Delta_{g}u\parallel_{g}
\end{equation*}
hence we have the inequality 
\begin{equation*}
\Vert D^{2}u\Vert\leq e^{\lambda_{M}}\Vert\Delta_{g}u\Vert_{g}+(1/\sqrt
{2})\Vert Du\Vert_{g}
\end{equation*}
which implies the given result 1.

2.\ Under the hypothesis H$_{c}$ we have 
\begin{equation*}
e^{\lambda_{M}}|\tau|\leq C.
\end{equation*}
the result 2 follows from the definitions of $\varepsilon$ and $\varepsilon
_{1}.$

\begin{lemma}
The following estimate holds if $\varepsilon_{v}\leq c$ (hypothesis H$_{c}$) 
\begin{equation*}
\frac{1}{2}e^{-2\omega}\parallel D|h|^{2}\parallel_{1}\leq C_{\sigma
}(\varepsilon^{2}+\varepsilon\varepsilon_{1})
\end{equation*}
\end{lemma}

Proof. We have: 
\begin{equation*}
\parallel D|h^{2}|\parallel_{1}\leq2\parallel h\parallel\parallel Dh\parallel
\end{equation*}
We have shown in a previous section that the $L^{2}$ norm of $h$ and $Dh$
can be estimated through the first and second energies. We have found 
\begin{equation*}
\parallel h\parallel\leq e^{\lambda_{M}}\varepsilon
\end{equation*}
and under the hypothesis H$_{c}$ 
\begin{equation*}
\parallel Dh\parallel\leq
e^{\lambda_{M}}\{\varepsilon+CC_{\sigma}\varepsilon(\varepsilon+%
\varepsilon_{1})\}
\end{equation*}
The given result follows from the bound of $e^{2(\lambda_{M}-\omega)}.$

We now estimate the last term in $Dh_{\omega},$ i.e. $\frac{1}{2}e^{-2\omega
}\parallel D|\overset{.}{u}|\parallel_{1}.$ We will use the following
estimates of $L^{4}$ norms of $Du,$ $u^{\prime}$ and $h:$

\begin{lemma}
1. Under the hypothesis H$_{c}$ the $L^{4}$ norms of $Du,$ $u^{\prime}$ and
h are estimated by:

\begin{equation*}
\Vert u^{\prime}\Vert_{4}^{2}\equiv\Vert\mid u^{\prime}\mid^{2}\Vert\leq
CC_{\sigma}\tau^{2}\{\varepsilon^{2}+\varepsilon\varepsilon_{1}\}
\end{equation*}
and 
\begin{equation*}
\Vert Du\Vert_{4}^{2}\equiv\Vert\mid Du\mid^{2}\Vert\leq CC_{\sigma
}\{\varepsilon^{2}+\varepsilon\varepsilon_{1}\}
\end{equation*}
2. The L$^{4}$ norm of h is estimated by 
\begin{equation*}
\Vert h\Vert_{4}^{2}\equiv\parallel|h|^{2}\parallel\leq CC_{\sigma
}\{e^{2\lambda_{M}}\varepsilon^{2}+C_{\sigma}e^{\lambda_{M}}\varepsilon
^{2}(\varepsilon+\varepsilon_{1})\}
\end{equation*}
\end{lemma}

Proof.

1. Immediate consequence of the inequalities proved in the final section on
local existence, and the definitions.

2. The inequality for the $L^{2}$ norm of $|h|^{2}$ is also proved through
the Sobolev inequality

\begin{equation*}
\parallel|h|^{2}\parallel\leq C_{\sigma}\{\parallel h\parallel^{2}+\parallel
D|h|^{2}\parallel_{1}\}
\end{equation*}
which gives, using previous results 
\begin{equation*}
\parallel|h|^{2}\parallel\leq C_{\sigma}e^{2\lambda_{M}}\{\varepsilon
^{2}+C\varepsilon^{2}(\varepsilon+\varepsilon_{1})\}.
\end{equation*}

\begin{lemma}
We have 
\begin{equation*}
\frac{1}{2}e^{-2\omega}\Vert D|\dot{u}|^{2}\Vert_{1}\leq CC_{\sigma
}(\varepsilon^{2}+\varepsilon\varepsilon_{1})
\end{equation*}
\end{lemma}

Proof. We have

\begin{equation*}
\Vert D|\dot{u}|^{2}\Vert_{1}\leq2\Vert\dot{u}\Vert\Vert D\dot{u}\Vert
\end{equation*}
Recall that 
\begin{equation*}
\dot{u}=e^{2\lambda}u^{\prime},u^{\prime}=N^{-1}\partial_{0}u,\text{ and }%
\mu_{\sigma}=e^{-2\lambda}\mu_{g}
\end{equation*}
hence 
\begin{equation*}
\parallel\overset{.}{u}\parallel\leq e^{\lambda_{M}}\parallel u^{\prime
}\parallel_{g}\leq e^{\lambda_{M}}\varepsilon
\end{equation*}
we have 
\begin{equation*}
D_{a}\dot{u}=e^{2\lambda}[D_{a}u^{\prime}+2u^{\prime}D_{a}\lambda]
\end{equation*}
\begin{equation*}
\Vert D\dot{u}\Vert^{2}=\int_{\Sigma}e^{4\lambda}\{|Du^{\prime}|^{2}+4|u^{%
\prime}|^{2}|D\lambda|^{2}+2D^{a}|u^{\prime}|^{2}D_{a}\lambda )\}\mu_{\sigma}
\end{equation*}
after integration by parts 
\begin{equation*}
\Vert D\dot{u}\Vert^{2}=\int_{\Sigma}e^{4\lambda}[|Du^{\prime}|^{2}-|u^{%
\prime}|^{2}(4|D\lambda|^{2}+2\Delta\lambda)]\mu_{\sigma}
\end{equation*}
When \ $R(\sigma)=-1$ \ we have 
\begin{equation*}
2\Delta\lambda=(\frac{1}{2}e^{2\lambda}\tau^{2}-1)-(e^{-2%
\lambda}|h|^{2}+|Du|^{2}+e^{2\lambda}|u^{\prime}|^{2})
\end{equation*}
It results from the estimate of $\lambda_{m}$ that on $\Sigma_{t}$ 
\begin{equation*}
\frac{1}{2}e^{2\lambda}\tau^{2}-1\geq0
\end{equation*}
hence 
\begin{equation*}
\parallel D\dot{u}\parallel^{2}\leq\int_{\Sigma}e^{4\lambda}\{|Du^{\prime
}|^{2}+|u^{\prime}|^{2}(|Du|^{2}+e^{-2\lambda}|h|^{2}+e^{2\lambda}|u^{\prime
}|^{2})\}\mu_{\sigma}
\end{equation*}
\TEXTsymbol{>}From which we deduce, using the Cauchy-Schwarz inequality 
\begin{equation*}
\Vert D\dot{u}\Vert^{2}\leq e^{4\lambda_{M}}\{\Vert Du^{\prime}\Vert
^{2}+\parallel|u^{\prime}|^{2}\parallel(\parallel|Du|^{2}\parallel
+e^{-2\lambda_{m}}\parallel|h|^{2}\parallel+e^{2\lambda_{M}}\parallel
|u^{\prime}|^{2}\parallel)\}
\end{equation*}
which we write, using previous results 
\begin{align*}
\Vert D\dot{u}\Vert^{2} & \leq
e^{4\lambda_{M}}\tau^{2}\{\varepsilon_{1}^{2}+CC_{\sigma}[\varepsilon^{2}+%
\varepsilon\varepsilon_{1})][(e^{2\lambda
_{M}}\tau^{2}+1)(\varepsilon^{2}+\varepsilon\varepsilon_{1})+ \\
&
+e^{-2\lambda_{m}}e^{2\lambda_{M}}(\varepsilon^{2}+CC_{\sigma}%
\varepsilon^{2}(\varepsilon+\varepsilon_{1})]\}
\end{align*}

Using once more the estimates resulting from the H hypothesis we obtain 
\begin{equation*}
\Vert D\dot{u}\Vert\leq Ce^{\lambda_{M}}\{\varepsilon_{1}+C_{\sigma
}[\varepsilon^{2}+\varepsilon\varepsilon_{1}]\}\leq
CC_{\sigma}e^{\lambda_{M}}(\varepsilon+\varepsilon_{1})
\end{equation*}
Assembling inequalities and the bound of $e^{2(\lambda_{M}-\omega)}$ leads
to the given result.

The following theorem is a straightforward consequence of our lemmas.

\begin{theorem}
There exists numbers $C$ and C$_{\sigma}$ such that the L$^{\infty}$ norm of
v is bounded by the following inequality 
\begin{equation}
\parallel v\parallel_{\infty}\equiv\varepsilon_{v}\leq
CC_{\sigma}(\varepsilon^{2}+\varepsilon\varepsilon_{1})
\end{equation}
\end{theorem}

Proof. Recall that there exists a Sobolev constant $C_{\sigma}$ such that

\begin{equation*}
\Vert v\Vert_{\infty}\leq C_{\sigma}\{\Vert D|Du|^{2}\Vert_{1}+e^{-2\omega
}(\Vert D|h|^{2}\Vert_{1}+\Vert D|\dot{u}|^{2}\Vert_{1})\}
\end{equation*}

The three terms in the sum have been evaluated in previous lemmas.

\subsection{Bound on derivatives.}

The equation satisfied by $\lambda$%
\begin{equation*}
\Delta\lambda=f(\lambda)\equiv\frac{1}{4}\tau^{2}e^{2\lambda}-\frac{1}{2}%
(\mid h\mid^{2}+\mid\dot{u}\mid^{2})e^{-2\lambda}-\frac{1}{2}(1+\mid
Du\mid^{2})
\end{equation*}
implies after multiplication by $\lambda-\overset{\_}{\lambda}$ and
integration on $\Sigma$%
\begin{equation*}
\parallel D\lambda\parallel^{2}\leq\parallel\lambda-\overset{\_}{\lambda }%
\parallel\parallel f(\lambda)\parallel
\end{equation*}
The Poincare inequality gives 
\begin{equation*}
\parallel\lambda-\overset{\_}{\lambda}\parallel\leq I_{\sigma}\parallel
D\lambda\parallel
\end{equation*}
therefore 
\begin{equation*}
\parallel D\lambda\parallel\leq I_{\sigma}\parallel f(\lambda)\parallel \leq%
\frac{1}{4}(\tau^{2}e^{2\lambda_{M}}-1)V_{\sigma}^{\frac{1}{2}}+\frac
{1}{2}\{\parallel|Du|^{2}\parallel+e^{-2\lambda_{m}}\parallel|h|^{2}%
\parallel+e^{2\lambda_{M}}\parallel|u^{\prime}|\parallel^{2}\}
\end{equation*}
while 
\begin{equation*}
\parallel D^{2}\lambda\parallel^{2}\equiv\parallel\Delta\lambda\parallel
^{2}+\frac{1}{2}\parallel D\lambda\parallel^{2}\leq(1+\frac{I_{\sigma}^{2}}{2%
})\parallel f(\lambda\parallel^{2}
\end{equation*}
The $L^{2}(\sigma)$ norm of $f(\lambda)$ is bounded by the following
quantity: 
\begin{equation*}
\parallel f(\lambda)\parallel\leq\frac{1}{4}(\tau^{2}e^{2\lambda_{M}}-2)V_{%
\sigma}^{\frac{1}{2}}+\frac{1}{2}\{\parallel|Du|^{2}\parallel
+e^{-2\lambda_{m}}\parallel|h|^{2}\parallel+e^{2\lambda_{M}}\parallel
|u^{\prime}|\parallel^{2}\}
\end{equation*}
Previous estimations show that 
\begin{equation}
\parallel f(\lambda)\parallel\leq CC_{\sigma}(\varepsilon^{2}+\varepsilon
\varepsilon_{1})
\end{equation}
which gives the following theorem.

\begin{theorem}
Under the hypothesis H$_{c}$ the $H_{1}$ norm of $D\lambda$ satisfies the
inequality 
\begin{equation}
\parallel D\lambda\parallel_{H_{1}}\leq
CC_{\sigma}(\varepsilon^{2}+\varepsilon\varepsilon_{1})
\end{equation}
\bigskip
\end{theorem}

\section{Estimates in $W_{s}^{p}.$}

\subsection{Estimates for h in $W_{2}^{p}$.\protect\bigskip}

The estimates of $h$ in $W_{2}^{p}$, with $1<p<2$ (for definiteness we will
choose $p=\frac{4}{3})$ will be obtained using estimates for the conformal
factor $\lambda$ which have been obtained by using the $H_{1}$ norm of $h.$

\begin{theorem}
Under the H hypothesis there exist positive numbers C(c) and C$_{\sigma}$
such that the $W_{2}^{p}$ norm of h, choosing to be specific $p=\frac{4}{3},$
is bounded by 
\begin{equation*}
\parallel h\parallel_{W_{2}^{p}}\leq CC_{\sigma}e^{\lambda_{M}}\{\varepsilon
+(\varepsilon+\varepsilon_{1})^{2}\}
\end{equation*}
\end{theorem}

\begin{corollary}
It holds that 
\begin{equation*}
|\tau|\parallel h\parallel_{\infty}\leq CC_{\sigma}\{\varepsilon
+(\varepsilon+\varepsilon_{1})^{2}\}
\end{equation*}
and that 
\begin{equation*}
\parallel h\parallel_{L^{\infty}(g)}\leq CC_{\sigma}|\tau|\{\varepsilon
+(\varepsilon+\varepsilon_{1})^{2}\}
\end{equation*}
\end{corollary}

Proof$.$ We recall that for any function f on a compact manifold one has, if 
$p\leq2,$%
\begin{equation*}
\parallel f\parallel_{p}\leq V_{\sigma}^{\frac{1}{p}-\frac{1}{2}}\parallel
f\parallel
\end{equation*}
We deduce therefore from the $H_{s}$ estimate of section 6.2 that ($C_{0}$
is a given number, $V_{\sigma}=|4\pi\chi|$ is a constant) 
\begin{equation*}
\parallel q\parallel_{W_{2}^{p}}\leq C_{0}\parallel q\parallel\leq
C_{0}\parallel h\parallel\leq\sqrt{2}C_{0}e^{\lambda_{M}}\varepsilon
\end{equation*}

To estimate $h$ in $W_{2}^{p}$ it remains to estimate $r$ in $W_{2}^{p}.$

It results from elliptic theory that on each $\Sigma_{t}$ the tensor $r$
satisfies for each $1<p<\infty$ the following estimate 
\begin{equation*}
\parallel r\parallel_{W_{2}^{p}}\leq C_{\sigma}\parallel Du.\overset{.}{u}%
\parallel_{W_{1}^{p}}
\end{equation*}
We choose 
\begin{equation*}
p=\frac{4}{3}
\end{equation*}
We have 
\begin{equation*}
\parallel Du.\overset{.}{u}\parallel_{\frac{4}{3}}\leq\parallel Du\parallel
\parallel\overset{.}{u}\parallel_{4}\leq e^{\lambda_{M}}\varepsilon
(\varepsilon^{2}+\varepsilon\varepsilon_{1})^{\frac{1}{2}}
\end{equation*}
because 
\begin{equation*}
\parallel Du\parallel=\parallel Du\parallel_{g}\leq\varepsilon
\end{equation*}
and, under the H$_{c}$ hypothesis 
\begin{equation*}
\parallel\overset{.}{u}\parallel_{4}\equiv\parallel|\overset{.}{u}%
|^{2}\parallel^{\frac{1}{2}}\leq e^{2\lambda_{M}}\Vert\mid u^{\prime}\mid
^{2}\Vert^{\frac{1}{2}}\leq CC_{\sigma}e^{\lambda_{M}}(\varepsilon
^{2}+\varepsilon\varepsilon_{1})^{\frac{1}{2}}
\end{equation*}
We now estimate 
\begin{equation*}
\parallel D(Du.\overset{.}{u})\parallel_{\frac{4}{3}}\leq\parallel
D^{2}u\parallel\parallel\overset{.}{u}\parallel_{4}+\parallel Du\parallel
_{4}\parallel D\overset{.}{u}\parallel
\end{equation*}
Using previous estimates we obtain by a straightforward calculation

\begin{equation*}
\parallel D(Du.\overset{.}{u})\parallel_{\frac{4}{3}}\leq CC_{\sigma
}e^{\lambda_{M}}\{\varepsilon^{\frac{1}{2}}(\varepsilon+\varepsilon _{1})^{%
\frac{3}{2}}+\varepsilon^{\frac{3}{2}}(\varepsilon+\varepsilon _{1})^{\frac{1%
}{2}}\}
\end{equation*}
The result of the theorem follows from the bound of $\varepsilon$ by $%
\varepsilon+\varepsilon_{1}$.

Proof of corollary.

1. The Sobolev embedding theorem, 
\begin{equation*}
\parallel h\parallel_{\infty}\leq C_{\sigma}\parallel h\parallel_{W_{2}^{p}}%
\text{ \ \ \ if \ \ \ }p>1,
\end{equation*}
and the estimate of $e^{\lambda_{M}}|\tau|.$

2. 
\begin{equation*}
\parallel
h\parallel_{L^{\infty}(g)}=Sup_{\Sigma}\{g^{ac}g^{bd}h_{ab}h_{cd}\}^{\frac{1%
}{2}}\leq e^{-2\lambda_{m}}\parallel h\parallel_{\infty}\leq\frac{1}{2}%
\tau^{2}\parallel h\parallel_{\infty}
\end{equation*}

\subsection{W$_{3}^{p}$ estimates for N.}

\subsubsection{$H_{2}$ estimates of N.}

\begin{theorem}
There exist numbers $C=C(c)$ and $C_{\sigma}$ such that the $H_{2}$norm of $%
N $ satisfies the inequality 
\begin{equation*}
\parallel2-N\parallel_{H_{2}}\leq CC_{\sigma}(\varepsilon^{2}+\varepsilon
\varepsilon_{1})
\end{equation*}
\end{theorem}

\begin{corollary}
The minimum $N_{m}$ of $N$ is such that 
\begin{equation*}
0\leq2-N_{m}\leq CC_{\sigma}(\varepsilon^{2}+\varepsilon\varepsilon_{1})
\end{equation*}
\bigskip
\end{corollary}

Proof. We write the equation satisfied by $N$ in the form

\begin{equation}
\Delta(2-N)-(2-N)=\beta
\end{equation}
with, having chosen the parameter t such that $\partial_{t}\tau=\tau^{2},$%
\begin{equation*}
\beta\equiv(2-N)(e^{2\lambda}\frac{1}{2}\tau^{2}-1)-N(e^{2\lambda}\mid
u^{\prime}\mid^{2}+e^{-2\lambda}\mid h\mid^{2})
\end{equation*}
The standard elliptic estimate applied to the form given to the lapse
equation gives 
\begin{equation}
\parallel2-N\parallel_{H_{2}}\leq C_{\sigma}\parallel\beta\parallel
\end{equation}
Since $0<N\leq2$ and $e^{-2\lambda}\leq\frac{1}{2}\tau^{2}$ it holds that 
\begin{equation}
\parallel\beta\parallel\leq2(\frac{1}{2}e^{2\lambda_{M}}\tau^{2}-1)V_{\sigma
}^{1/2}+2(e^{2\lambda_{M}}\parallel|u^{\prime}|^{2}\parallel+\frac{1}{2}%
\tau^{2}\parallel|h|^{2}\parallel)
\end{equation}
The $L^{4}$ norms of $h$ and $u^{\prime}$ as well as $\frac{1}{2}%
e^{2\lambda_{M}}\tau^{2}-1$ have been estimated in the section conformal
factor estimate. We deduce from these estimates the bound 
\begin{equation*}
\parallel\beta\parallel\leq CC_{\sigma}(\varepsilon^{2}+\varepsilon
\varepsilon_{1}).
\end{equation*}
which gives the result of the theorem.

The corollary is a consequence of the Sobolev embedding theorem.

\begin{theorem}
Under the hypothesis H$_{c}$ there exist numbers $C$ depending only on $c$
and $C_{\sigma}$ such that if $1<p<2$, for instance $p=\frac{4}{3}$ 
\begin{equation*}
\varepsilon_{DN}\equiv\parallel2-N\parallel_{W_{3}^{p}}\leq CC_{\sigma
}(\varepsilon^{2}+\varepsilon\varepsilon_{1}).
\end{equation*}

\begin{corollary}
The gradient of $N$ satisfies the inequality: 
\begin{equation*}
\parallel DN\parallel_{L^{\infty}(g)}|\leq CC_{\sigma}|\tau|(\varepsilon
^{2}+\varepsilon\varepsilon_{1})
\end{equation*}
\end{corollary}
\end{theorem}

Proof. We have 
\begin{equation}
|\beta|\leq(2-N_{m})(\frac{1}{2}e^{2\lambda_{M}}\tau^{2}-1)+2(e^{2%
\lambda_{M}}|u^{\prime}|^{2}+\frac{1}{2}\tau^{2}|h|^{2})
\end{equation}
We apply the standard elliptic estimate 
\begin{equation}
\parallel2-N\parallel_{W_{s+2}^{p}}\leq C_{\sigma}\parallel\beta
\parallel_{W_{s}^{p}}
\end{equation}
with now $1<p<2,s=1.$

We have for any $p\leq2,$%
\begin{equation*}
\parallel\beta\parallel_{p}\leq V_{\sigma}^{\frac{1}{p}-\frac{1}{2}%
}\parallel\beta\parallel
\end{equation*}
We have already estimated $\parallel\beta\parallel.$

To estimate $\parallel\beta\parallel_{W_{1}^{p}}$ we compute 
\begin{align*}
D\beta & \equiv\lbrack(2-N)e^{2\lambda}\tau^{2}-2N(e^{2\lambda}\mid
u^{\prime}\mid^{2}-e^{-2\lambda}\mid h\mid^{2})]D\lambda \\
-DN[\frac{1}{2}e^{2\lambda}\tau^{2}-1-e^{2\lambda} & \mid u^{\prime}\mid
^{2}-e^{-2\lambda}\mid h\mid^{2}]-N[e^{2\lambda}D\mid u^{\prime}\mid
^{2}+e^{-2\lambda}D\mid h\mid^{2}]
\end{align*}
We have therefore, with $\frac{1}{q}+\frac{1}{q^{\prime}}=\frac{1}{p}$, and
using estimates obtained for $\lambda$ under the H hypothesis 
\begin{align*}
& \parallel D\beta\parallel_{p}\leq CC_{\sigma}\{(2-N_{m})\parallel
D\lambda\parallel_{p}+(\varepsilon^{2}+\varepsilon\varepsilon_{1})\parallel
DN\parallel_{p}+ \\
\lbrack e^{2\lambda_{M}} & \parallel|u^{\prime}|^{2}\parallel_{q^{\prime}}+%
\frac{1}{2}\tau^{2}\parallel|h|^{2}\parallel_{q^{\prime}}][4\parallel
D\lambda\parallel_{q}+\parallel DN\parallel_{q}]+A\}
\end{align*}
with 
\begin{equation*}
A\equiv2[e^{2\lambda_{M}}\parallel D\mid u^{\prime}\mid^{2}\parallel_{p}+%
\frac{1}{2}\tau^{2}\parallel D\mid h\mid^{2}\parallel_{p}]
\end{equation*}
To bound the first line we recall that the $L^{p}$ norms of $D\lambda$ and $%
DN$ are bounded by their $L^{2}$ norms estimated before. To estimate the
second line (except for $A$) we choose $p=\frac{4}{3},q=4,q^{\prime}=2.$ We
find quantities bounded before and the $L^{4}$ norm of $D\lambda$ and $DN$
which can be estimated in terms of their $H_{1}$ norms bounded before.

To bound $A$ we write again, with $p=\frac{4}{3}$: 
\begin{equation*}
\parallel D|u^{\prime}|^{2}\parallel_{p}\leq2\parallel u^{\prime}\parallel
_{4}\parallel Du^{\prime}\parallel,\text{ since }\frac{1}{4}+\frac{1}{2}=%
\frac{1}{p}
\end{equation*}
This inequality and corresponding estimates for $h$ give: 
\begin{equation*}
A\leq CC_{\sigma}(\varepsilon^{2}+\varepsilon\varepsilon_{1})
\end{equation*}
The $H_{1}$ bound found above for $DN$ and $D\lambda$ permits the obtention
of the given result.

The corollary is a consequence of the Sobolev embedding theorem and the
relation between $\sigma$ and $g$ norms: 
\begin{equation*}
\parallel DN\parallel_{L^{\infty}(g)}\leq e^{-\lambda_{m}}\parallel
DN\parallel_{\infty}\leq e^{-\lambda_{m}}C_{\sigma}\parallel DN\parallel
_{W_{2}^{p}}\leq
CC_{\sigma}|\tau|(\varepsilon^{2}+\varepsilon\varepsilon_{1})
\end{equation*}

\section{Corrected energy estimates.}

We have obtained in section 6 a bound for the first energy and a decay for
the second energy. These bounds prove unsufficient to control the behaviour
in time of the Teichmuller parameters. The right hand side of the first
energy inequality is non positive, as well as the quadratic term of the
right hand side of the second energy inequality, but the space derivatives
are lacking in those right hand sides which would make them negative
definite. The introduction of corrected energies enables one to obtain such
a definiteness, compensating some terms by others, and leading to better
decay estimates.

\subsection{Corrected first energy.}

\subsubsection{Definition and lower bound.}

One defines as follows a corrected first energy where $\alpha$ is a
constant, which we will choose positive:

\begin{equation}
E_{\alpha}(t)=E(t)-\alpha\tau\int_{\Sigma_{t}}(u-\overset{\_}{u}).u^{\prime
}\mu_{g}
\end{equation}
where we have denoted by $\overset{\_\text{ }}{u}$ the mean value of $u$, a
scalar function, on $\Sigma_{t}:$

\begin{equation*}
\overset{\_}{u}=\frac{1}{Vol_{\sigma }\Sigma _{t}}\int_{\Sigma _{t}}u\mu
_{\sigma }
\end{equation*}
An estimate of $E_{\alpha }$ will give estimates of the $L^{2}$ norms of the
derivatives of $u$ and of $h$ if there exists a $K>0,$ independent of $t,$
such that

\begin{equation}
E(t)\leq KE_{\alpha}(t)\equiv K[E(t)-\alpha\tau\int_{\Sigma_{t}}(u-\overset{%
\_}{u}).u^{\prime}\mu_{g}]
\end{equation}
We set 
\begin{equation}
I_{0}\equiv\frac{1}{2}|u^{\prime}|^{2},\text{ and }I_{1}\equiv\frac{1}{2}%
|Du|_{g}^{2}
\end{equation}
and 
\begin{equation*}
x_{0}=\int_{\Sigma_{t}}I_{0}\mu_{g}\equiv\frac{1}{2}\parallel u^{\prime
}\parallel_{g}^{2},\text{ and }x_{1}=\frac{1}{2}\parallel Du\parallel_{g}^{2}
\end{equation*}

\noindent We estimate the complementary term through the Cauchy-Schwarz
inequality

\begin{equation*}
|\int_{\Sigma_{t}}(u-\overset{\_}{u}).u^{\prime}\mu_{g}|\leq||u-\overset{\_}{%
u}||_{g}||u^{\prime}||_{g}.
\end{equation*}

\noindent We will use the Poincar\'{e} inequality on the compact manifold ($%
\Sigma,\sigma)$ to estimate the $L^{2}(\sigma)$ norm of $u-\bar{u}$: 
\begin{equation}
||u-\overset{\_}{u}||_{g}\leq e^{\lambda_{M}}||u-\overset{-}{u}||\leq
e^{\lambda_{M}}\Lambda_{\sigma}^{-1/2}||Du||
\end{equation}
where $\parallel.\parallel$ denotes the $L^{2}$ norm on $\Sigma$ in the
metric $\sigma,\lambda_{M}$ is an upper bound of the conformal factor $%
\lambda$ and $\Lambda_{\sigma}$ is the first positive eigenvalue of the
operator $-\Delta\equiv-\Delta_{\sigma}$ acting on functions with mean value
zero. Note that $||Du||=||Du||_{g}$.

The inequality (27) to satisfy is implied by the two following ones: 
\begin{equation}
K\geq1
\end{equation}
and (to be satisfied by all $x_{0},x_{1}\geq0)$%
\begin{equation}
(K-1)(x_{0}+x_{1})-2|\alpha\tau|Ke^{\lambda_{M}}\Lambda_{\sigma}^{-\frac{1}{2%
}}x_{0}^{\frac{1}{2}}x_{1}^{\frac{1}{2}}\geq0
\end{equation}
this quadratic form in the $x$'s will be always non negative if $K\geq1$ and
its discriminant is non positive. This last condition reads 
\begin{equation*}
aK\leq K-1
\end{equation*}
with 
\begin{equation}
a\equiv\frac{\alpha|\tau|e^{\lambda_{M}}}{\Lambda_{\sigma}^{\frac{1}{2}}}
\end{equation}
A necessary and sufficient condition for the existence of $K\geq1$ and $K$
finite is therefore 
\begin{equation}
a^{2}\equiv\alpha^{2}\tau^{2}e^{2\lambda_{M}}\Lambda_{\sigma}^{-1}<1
\end{equation}
Any $K$ such that 
\begin{equation}
K\geq\frac{1}{1-a}
\end{equation}
satisfies then the required conditions.

It is known that given a 2-manifold $\Sigma$ of genus $G>1$ there is an open
subset of Teichmuller space such that for metrics $\sigma\in M_{-1}$
projecting on this open set it holds 
\begin{equation}
8\Lambda_{\sigma}=1+\delta_{\sigma},\text{ with }\delta_{\sigma}>0
\end{equation}
We now choose 
\begin{equation*}
\alpha=\frac{1}{4}
\end{equation*}
The condition $a<1$ then reads 
\begin{equation}
(\frac{\tau^{2}e^{2\lambda_{M}}}{2})(\frac{1}{1+\delta_{\sigma}})<1,\text{ }
\end{equation}
that is using estimates on the conformal factor 
\begin{equation*}
\text{ }C(\varepsilon^{2}+C_{\sigma}\varepsilon\varepsilon_{1})<\delta
_{\sigma}
\end{equation*}

\subsubsection{Time derivative of the corrected energy.}

We set:

\begin{equation*}
\frac{dE_{\alpha}}{dt}=\frac{dE}{dt}-R_{\alpha}
\end{equation*}

\noindent with (the terms explicitly containing the shift $\nu$ give an
exact divergence which integrates to zero) 
\begin{equation*}
R_{\alpha}=\alpha\tau\int_{\Sigma_{t}}\{\partial_{0}u^{\prime}.(u-\overset{\_%
}{u})+u^{\prime}.\partial_{0}(u-\overset{\_}{u})-N\tau u^{\prime }.(u-%
\overset{\_}{u})\}\mu_{g}
\end{equation*}

\begin{equation}
+\alpha\frac{d\tau}{dt}\int_{\Sigma_{t}}u^{\prime}.(u-\overset{\_}{u})\mu_{g}
\end{equation}

\noindent The function $\gamma\equiv u$ satisfies the wave equation 
\begin{equation}
-N^{-1}\partial_{0}(N^{-1}\partial_{0}u)+N^{-1}\nabla^{a}(N\partial
_{a}u)+N^{-1}\tau\partial_{0}u=0
\end{equation}

\noindent Some elementary computations and integration by parts show that 
\begin{equation*}
R_{\alpha}=\alpha\tau\int_{\Sigma_{t}}\{|u^{\prime}|^{2}-|Du|_{g}^{2}\}N%
\mu_{g}
\end{equation*}
\begin{equation*}
-\alpha\tau\int_{\Sigma_{t}}u^{\prime}.\partial_{t}\overset{\_}{u}\mu
_{g}+\alpha\frac{d\tau}{dt}\int_{\Sigma_{t}}(u-\overset{\_}{u}%
).u^{\prime}\mu_{g}
\end{equation*}

\textbf{Lemma. }If $u$ satisfies the wave equation the quantity 
\begin{equation*}
\int_{\Sigma_{t}}u^{\prime}\mu_{g}
\end{equation*}
is conserved in time

Proof. Integration on ($\Sigma_{t},g)$ of the wave equation (multiplied by $%
N $) shows that on a compact manifold, where exact divergences integrate to
zero, one has 
\begin{equation*}
\frac{d}{dt}\int_{\Sigma_{t}}u^{\prime}\mu_{g}=\int_{\Sigma_{t}}(\partial
_{0}u^{\prime}-N\tau u^{\prime})\mu_{g}=0
\end{equation*}

To simplify the proofs we will suppose in all that follows that 
\begin{equation}
\int_{\Sigma_{t}}u^{\prime}\mu_{g}=0
\end{equation}
Then $R_{a\text{ }}$reduces to, since $\partial_{t}\overset{\_}{u}$ is
constant on $\Sigma_{t},$ 
\begin{equation}
R_{\alpha}=\alpha\tau\int_{\Sigma_{t}}\{|u^{\prime}|^{2}-|Du|_{g}^{2}\}N%
\mu_{g}+\alpha\frac{d\tau}{dt}\int_{\Sigma_{t}}(u-\overset{\_}{u}%
).u^{\prime}\mu_{g}
\end{equation}

\subsubsection{Decay of the corrected first energy.}

In the corrected energy inequality we have seen appear the quantity $%
d\tau/dt.$ To obtain a differential inequality we have to make a choice of $%
\tau$ as a function of $t.$ We wish to work in the expanding direction of
our spacetime, where $\tau,$ with our sign convention for the extrinsic
curvature, starts from a negative value $\tau_{0}$ and increases, eventually
up to the moment of maximum expansion where $\tau=0.$ \ We have made
(section 5, notations) the choice 
\begin{equation}
\tau=-t^{-1\text{ }},\text{ \ \ \ }t\in\lbrack t_{0},\infty),\text{ \ \ }%
t_{0}>0,\text{ \ \ }\frac{d\tau}{dt}=\frac{1}{t^{2}}=\tau^{2}.
\end{equation}

We obtain, using the value of dE/dt and R$_{\alpha},$ that

\begin{equation}
\frac{dE_{\alpha}}{dt}=\tau\int_{\Sigma_{t}}\{[\frac{1}{2}|h|^{2}+(\frac{1}{2%
}-\alpha)|u^{\prime}|^{2}+\alpha|Du|_{g}^{2}]N-\alpha\tau u^{\prime }.(u-%
\overset{\_}{u})\}\mu_{g}
\end{equation}
we look for a positive number k such that the difference 
\begin{equation*}
\frac{dE_{\alpha}}{dt}-k\tau E_{\alpha}
\end{equation*}
can be estimated with higher order terms. We choose 
\begin{equation*}
\alpha=\frac{1}{4},\allowbreak k=1
\end{equation*}
We have then 
\begin{equation*}
\frac{dE_{1/4}}{dt}-\tau E_{1/4}=\tau\int_{\Sigma_{t}}\{\frac{1}{2}%
|h|_{g}^{2}(N-1)+[\frac{1}{2}N-1](I_{0}+I_{1})\}\mu_{g}
\end{equation*}

Which we write

\begin{equation*}
\frac{dE_{1/4}}{dt}-\tau E_{1/4}=\tau\int_{\Sigma_{t}}\{\frac{1}{2}%
|h|_{g}^{2}(1+N-2)+\frac{1}{2}(N-2)(I_{0}+I_{1})\}\mu_{g}
\end{equation*}
The right hand side is the sum of a negative term and a term which can be
considered as a non linear term in the energies because we have proved that
(cf section 9.2 on $N$ estimates):

\begin{equation*}
0\leq2-N\leq2-N_{m}\leq CC_{\sigma}(\varepsilon^{2}+\varepsilon\varepsilon
_{1})
\end{equation*}
Therefore we obtain the following theorem (remember that $\tau<0)$:

\begin{theorem}
The corrected first energy with $\alpha=\frac{1}{4}$ satisfies the
differential equation
\end{theorem}

\begin{equation}
\frac{dE_{1/4}}{dt}=\tau E_{1/4}+|\tau|A\text{ \ \ \ \ with \ \ \ }A\leq
CC_{\sigma}\varepsilon^{2}(\varepsilon^{2}+\varepsilon\varepsilon_{1})
\end{equation}

\subsection{Corrected second energy.}

\subsubsection{Definition and lowerbound.}

We define a corrected second energy $E_{\alpha}$ by the formula, with $%
\alpha $ some constant 
\begin{equation*}
E_{\alpha}^{(1)}(t)=E^{(1)}+C_{\alpha}
\end{equation*}
with 
\begin{equation*}
C_{\alpha}=\alpha\tau\int_{\Sigma_{t}}\Delta_{g}u.u^{\prime}\mu_{g}
\end{equation*}

This corrected second energy will give bounds on the derivatives of $Du$ and 
$u^{\prime}$ if there exists a number $K$ \TEXTsymbol{>}0 such that: 
\begin{equation}
E^{(1)}\leq KE_{\alpha}^{(1)}
\end{equation}
The hypothesis $\bar{u}^{\prime}=0$ is not necessary here because on a
compact manifold $\int_{\Sigma_{t}}\Delta_{g}u.\overset{\_}{u}%
^{\prime}\mu_{g}=0.$ We obtain the estimate, analogous to one obtained in
the previous section, 
\begin{equation*}
\int_{\Sigma_{t}}\Delta_{g}u.u^{\prime}\mu_{g}\leq\parallel\Delta
_{g}u\parallel_{g}\parallel u^{\prime}-\overset{\_}{u}^{\prime}\parallel
_{g}\leq\parallel\Delta_{g}u\parallel_{g}e^{\lambda_{M}}\Lambda_{\sigma }^{-%
\frac{1}{2}}\parallel Du^{\prime}\parallel_{g}
\end{equation*}
The same $\ K$ as in the previous section satisfies the required inequality
when we choose $\alpha=\frac{1}{4}.$

\subsubsection{Time derivative of the corrected second energy.}

We have

\begin{equation*}
dC_{\alpha}/dt=\alpha\tau\int_{\Sigma_{t}}[\partial_{0}\Delta_{g}u.u^{\prime
}+\Delta_{g}u.\partial_{0}u^{\prime}-N\tau\Delta_{g}u.u^{\prime}]\mu
_{g}+\alpha\frac{d\tau}{dt}\int_{\Sigma_{t}}\Delta_{g}u.u^{\prime}\mu_{g}
\end{equation*}
We recall that (indices are raised with g in the next few lines) 
\begin{equation*}
\partial_{0}\Delta_{g}u=\Delta_{g}(Nu^{\prime})+N\tau%
\Delta_{g}u+2Nh_{g}^{ab}\nabla_{a}\partial_{b}u+\partial_{c}u[2%
\nabla_{a}(Nk^{ac})-\tau \partial^{c}N]
\end{equation*}
Partial integration together with the splitting $k_{ab}=h_{ab}+\frac{1}{2}%
g_{ab}\tau$ , and the equation 
\begin{equation*}
^{(3)}R_{0}^{c}\equiv-N\nabla_{a}k^{ac}=\partial_{0}u.\partial^{c}u
\end{equation*}
gives: 
\begin{align*}
\int_{\Sigma_{t}}\partial_{0}\Delta_{g}u.u^{\prime}\mu_{g} & =\int
_{\Sigma_{t}}\{-N|Du^{\prime}|_{g}^{2}-\partial^{a}N\partial_{a}u^{\prime
}.u^{\prime}+N\tau\Delta_{g}u.u^{\prime}+ \\
& 2Nh_{g}^{ab}\nabla_{a}\partial_{b}u.u^{\prime}+2u^{\prime}.\partial
_{c}u(\partial_{a}Nh_{g}^{ac}-\partial^{c}u.u^{\prime})\}\mu_{g}
\end{align*}

On the other hand and if $u$ satisfies the wave equation we find 
\begin{equation*}
\int_{\Sigma_{t}}\Delta_{g}u.\partial_{o}u^{\prime}\mu_{g}=\int_{\Sigma_{t}}%
\{N|\Delta_{g}u|^{2}+\partial^{a}N\partial_{a}u.\Delta_{g}u+N\tau u^{\prime
}.\Delta_{g}u\}\mu_{g}
\end{equation*}

These equalities give, if we make the choice $\tau=\frac{-1}{t}$, hence $%
\frac{d\tau}{dt}=\tau^{2}:$ 
\begin{align*}
\frac{dC_{\alpha}}{dt} &
=\alpha\tau\int_{\Sigma_{t}}\{-N|Du^{\prime}|^{2}+N|\Delta_{g}u|^{2}+%
\partial^{a}N(\partial_{a}u.\Delta_{g}u+u^{\prime }.\partial_{a}u^{\prime})
\\
& +2Nh_{g}^{ab}\nabla_{a}\partial_{b}u.u^{\prime}+2u^{\prime}.\partial
_{c}u(\partial_{a}Nh_{g}^{ac}-u^{\prime}.\partial^{c}u)+(N+1)\tau\Delta
_{g}u.u^{\prime}\}\mu_{g}
\end{align*}
We have found an equality of the form 
\begin{equation*}
\frac{dE_{\alpha}^{(1)}}{dt}\equiv\frac{dE^{(1)}}{dt}+\frac{dC_{\alpha}}{dt}%
=\int_{\Sigma_{t}}\{\tau P_{\alpha}+\alpha\tau Q\}\mu_{g}+Z
\end{equation*}
with 
\begin{equation*}
P_{a}=N[2(1-\alpha)J_{0}+(1+2\alpha)J_{1}]+(N+1)\alpha\tau\Delta
_{g}u.u^{\prime}
\end{equation*}
and 
\begin{align*}
Q & \equiv\partial^{a}N(\partial_{a}u.\Delta_{g}u+u^{\prime}.\partial
_{a}u^{\prime}) \\
& \ +2Nh_{g}^{ab}\nabla_{a}\partial_{b}u.u^{\prime}+2u^{\prime}.\partial
_{c}u(\partial_{a}Nh_{g}^{ac}-u^{\prime}.\partial^{c}u)
\end{align*}
We see that $Q$ contains also terms only quadratic in the first and second
derivatives of $u$, but its integral will be bounded by non linear terms in
the energies through previous estimates on $DN$ and $h$.

We choose $\alpha=\frac{1}{4}.$ We split the integral of $P_{1/4}$ into
linear and non linear terms in the energies by writing 
\begin{equation*}
\int_{\Sigma_{t}}P_{1/4}\mu_{g}=3\int_{\Sigma_{t}}(J_{0}+J_{1}+\frac{1}{4}%
\tau\Delta_{g}u.u^{\prime})\mu_{g}+U\equiv3E_{1/4}^{(1)}+U
\end{equation*}
with non linear terms $U$ given by 
\begin{equation*}
U=\int_{\Sigma_{t}}(N-2)[\frac{3}{2}(J_{0}+J_{1})+\frac{1}{4}\tau\Delta
_{g}u.u^{\prime}]\mu_{g}
\end{equation*}

We are ready to prove the following theorem

\begin{theorem}
\textbf{\ }With the choice $\alpha=\frac{1}{4}$ and $\tau=-\frac{1}{t}$, $%
t>0,$ the corrected second energy satisfies the inequality 
\begin{equation*}
\frac{dE_{1/4}^{(1)}}{dt}=3\tau E_{1/4}^{(1)}+|\tau|^{3}B
\end{equation*}
where $B$ a polynomial in $\varepsilon$ and $\varepsilon_{1}$ with all terms
of order at least 3 and coefficients of the form $CC_{\sigma}.$
\end{theorem}

Proof. We have shown that 
\begin{equation*}
\frac{dE_{1/4}^{(1)}}{dt}=3\tau E_{1/4}^{(1)}+Z+\frac{1}{4}\tau\int Q\mu
_{g}+\tau U
\end{equation*}

We will estimate the various terms in the right hand side.

We obtain, using the bound of $2-N$ and the definition of $\varepsilon_{1}$ 
\begin{equation*}
|\tau U|\leq CC_{\sigma}|\tau|^{3}(\varepsilon^{2}+\varepsilon\varepsilon
_{1})(\varepsilon_{1}^{2}+\varepsilon\varepsilon_{1})
\end{equation*}
We now estimate $\int\tau Q\mu_{g},$ using its expression and the estimates
(cf.section 9) 
\begin{equation*}
\parallel h\parallel_{L^{\infty}(g)}\leq|\tau|\varepsilon_{h},\text{ with }%
\varepsilon_{h}=CC_{\sigma}\{\varepsilon+\varepsilon^{1/2}(\varepsilon
+\varepsilon_{1})^{3/2}\},
\end{equation*}
\begin{equation*}
\parallel DN\parallel_{L^{\infty}(g)}\leq|\tau|C_{\sigma}\varepsilon _{DN},%
\text{ with \ }\varepsilon_{DN}=CC_{\sigma}(\varepsilon^{2}+\varepsilon%
\varepsilon_{1})
\end{equation*}
we have, with $C_{0}$ a fixed number

\begin{align*}
& |\tau\int_{\Sigma_{t}}\{\partial^{a}N(\partial_{a}u.\Delta_{g}u+u^{\prime
}.\partial_{a}u^{\prime})+2u^{\prime}.\partial_{c}u(\partial_{a}Np^{ac}\}%
\mu_{g}| \\
& \ \leq C_{0}|\tau|^{3}(\varepsilon_{DN}+\varepsilon_{h})\varepsilon
\varepsilon_{1}+\varepsilon_{DN}\varepsilon_{h}\varepsilon^{2}
\end{align*}
while 
\begin{equation*}
|\tau\int_{\Sigma_{t}}2Np^{ab}\nabla_{a}\partial_{b}u.u^{\prime}\mu_{g}|%
\leq4\tau^{2}\varepsilon_{h}\varepsilon\parallel\nabla^{2}u\parallel_{g}
\end{equation*}
It holds on a 2 dimensional compact manifold 
\begin{equation*}
\parallel\nabla^{2}u\parallel_{g}^{2}=\parallel\Delta_{g}u\parallel_{g}^{2}-%
\frac{1}{2}\int_{\Sigma_{t}}R(g)|Du|_{g}^{2}\mu_{g}
\end{equation*}
Recall that 
\begin{equation*}
R(g)=-\frac{1}{2}\tau^{2}+|p|_{g}^{2}+|u^{\prime}|^{2}+|Du|_{g}^{2},\text{ \
\ with \ \ }|Du|_{g}^{2}\leq\frac{\tau^{2}}{2}|Du|^{2}
\end{equation*}
and 
\begin{equation*}
\int_{\Sigma_{t}}R(g)|Du|_{g}^{2}\mu_{g}=\int_{\Sigma_{t}}R(g)|Du|^{2}\mu_{%
\sigma}
\end{equation*}
therefore 
\begin{equation*}
\parallel\nabla^{2}u\parallel_{g}^{2}\leq
C_{0}\tau^{2}[\varepsilon_{1}^{2}+(1+\varepsilon_{h})\varepsilon^{2}]+[%
\parallel|u^{\prime}|^{2}\parallel+\tau^{2}\parallel|Du|^{2}\parallel]%
\parallel|Du|^{2}\parallel
\end{equation*}
The bounds on $L^{4}$ norms of $u^{\prime}$ and $Du$ give 
\begin{equation*}
\parallel\nabla^{2}u\parallel_{g}\leq|\tau|\varepsilon_{\nabla^{2}u},\text{
\ \ \ \ \ }\varepsilon_{\nabla^{2}u}=C_{0}\{\varepsilon_{1}^{2}+(1+%
\varepsilon
_{h})\varepsilon^{2}+CC_{\sigma}\varepsilon^{2}[\varepsilon_{1}+\varepsilon
]^{2}\}^{\frac{1}{2}}
\end{equation*}
Finally 
\begin{equation*}
|\tau\int_{\Sigma_{t}}\{2(u^{\prime}.\partial_{c}u)(u^{\prime}.\partial
^{c}u)\}\mu_{g}|\leq2|\tau|\parallel
u^{\prime}\parallel_{L^{4}(g)}^{2}\parallel Du\parallel_{L^{4}(g)}^{2}
\end{equation*}
We have, using previous estimates, 
\begin{equation*}
\parallel u^{\prime}\parallel_{L^{4}(g)}^{2}\leq e^{\lambda_{M}}\parallel
u^{\prime}\parallel_{4}\leq CC_{\sigma}e^{\lambda_{M}}\tau^{2}(\varepsilon
^{2}+\varepsilon\varepsilon_{1})
\end{equation*}
hence 
\begin{equation*}
\parallel u^{\prime}\parallel_{L^{4}(g)}^{2}\leq CC_{\sigma}|\tau
|(\varepsilon^{2}+\varepsilon\varepsilon_{1})
\end{equation*}
An inequality of the same type holds for $\parallel Du\parallel_{L^{4}(g)}.$

The estimate of $|\int\tau Q\mu_{g}|$ by the product of $|\tau|^{3}$ with
higher than 2 powers of the $\varepsilon^{\prime}s$ follows.

We now estimate Z. We recall that 
\begin{align}
Z & \equiv\int_{\Sigma_{t}}\{Np^{ab}\partial_{a}u^{\prime}.\partial
_{b}u^{\prime}+2Np^{ab}\nabla_{a}\partial_{b}u.\Delta_{g}u+(\nabla
_{b}(\partial^{a}N\partial_{a}u)+\tau\partial_{b}Nu^{\prime}).(\partial
^{b}u^{\prime})\}\mu_{g}  \notag \\
& +Y_{1}
\end{align}
Previous estimates give 
\begin{equation}
|Z|\leq|\tau|^{3}\{C_{0}\varepsilon_{h}\varepsilon_{1}^{2}+\varepsilon
_{DN}(\varepsilon_{1}^{2}+\varepsilon\varepsilon_{1})+4\varepsilon
_{1}\varepsilon_{\nabla^{2}u}\}+Y_{2}+|Y_{1}|
\end{equation}
with 
\begin{equation*}
Y_{2}\equiv|\int_{\Sigma_{t}}\{(\nabla_{b}\partial^{a}N)\partial
_{a}u.(\partial^{b}u^{\prime})\}\mu_{g}|
\end{equation*}
To bound $Y_{2}$ we use the $L^{4}$ norm of $\nabla^{2}N$ estimated in terms
of its $W_{3}^{p}$ norm in the section on lapse estimates$.$ Indeed 
\begin{equation*}
Y_{2}\leq|\tau|\varepsilon_{1}\parallel\nabla^{2}N\parallel_{L^{4}(g)}%
\parallel Du\parallel_{L^{4}(g)}
\end{equation*}
We have 
\begin{equation*}
|\nabla^{2}N|_{g}=e^{-2\lambda}|\nabla^{2}N|
\end{equation*}
hence 
\begin{equation*}
\parallel\nabla^{2}N\parallel_{L^{4}(g)}\leq e^{-\frac{3}{2}%
\lambda_{m}}\parallel\nabla^{2}N\parallel_{4}\leq C_{0}|\tau|^{\frac{3}{2}%
}\parallel\nabla^{2}N\parallel_{4}
\end{equation*}
On the other hand we recall the identity 
\begin{equation*}
\nabla_{a}\partial_{b}N\equiv D_{a}\partial_{b}N+\sigma^{cd}\partial
_{c}N\partial_{d}\lambda-\delta_{a}^{c}\partial_{b}\lambda\partial_{c}N-%
\delta_{b}^{c}\partial_{a}\lambda\partial_{b}N
\end{equation*}
By the Sobolev embedding theorem, with $p=\frac{4}{3}$ 
\begin{equation*}
\parallel D^{2}N\parallel_{4}\leq C_{\sigma}\parallel D^{2}N\parallel
_{W_{1}^{p}}\leq C_{\sigma}\varepsilon_{DN}
\end{equation*}
We also bound 
\begin{equation*}
\parallel D\lambda\parallel_{4}\leq C_{\sigma}\parallel D\lambda
\parallel_{H_{1}}
\end{equation*}
with 
\begin{equation*}
\parallel D\lambda\parallel_{H_{1}}\leq
CC_{\sigma}(\varepsilon^{2}+\varepsilon\varepsilon_{1})
\end{equation*}
and we obtain 
\begin{equation*}
\parallel\nabla^{2}N\parallel_{4}\leq
C_{\sigma}\varepsilon_{DN}(1+C(\varepsilon^{2}+\varepsilon\varepsilon_{1})
\end{equation*}
\bigskip Recall that 
\begin{equation*}
\parallel Du\parallel_{L^{4}(g)}\leq C_{0}|\tau|^{\frac{1}{2}}\parallel
Du\parallel_{4}\leq CC_{\sigma}|\tau|^{\frac{1}{2}}(\varepsilon+\varepsilon
^{\frac{1}{2}}\varepsilon_{1}^{\frac{1}{2}})
\end{equation*}
Finally 
\begin{equation*}
Y_{2}\leq
CC_{\sigma}|\tau|^{3}\varepsilon_{DN}[1+C(\varepsilon^{2}+\varepsilon%
\varepsilon_{1})]\varepsilon_{1}(\varepsilon+\varepsilon^{\frac
{1}{2}}\varepsilon_{1}^{\frac{1}{2}})
\end{equation*}
Recall that 
\begin{equation}
Y_{1}=\int_{\Sigma_{t}}\{(2\partial_{a}Np^{ac}-2N\partial^{c}u.u^{\prime
})\partial_{c}u+2\partial^{a}N\partial_{a}u^{\prime}+u^{\prime}\Delta
_{g}N\}.\Delta_{g}u\mu_{g}
\end{equation}
hence 
\begin{equation}
|Y_{1}|\leq|\tau|^{3}CC_{\sigma}\{\varepsilon_{DN}\varepsilon_{h}\varepsilon%
\varepsilon_{1}+\varepsilon_{DN}\varepsilon_{1}^{2}\}+Y_{3}+Y_{4}
\end{equation}
with 
\begin{equation*}
Y_{3}=|\int_{\Sigma_{t}}\{(-2N\partial^{c}u.u^{\prime})\partial_{c}u\}.%
\Delta_{g}u\mu_{g}|
\end{equation*}
\bigskip The term $Y_{3}$ can be estimated using the Holder inequality, 
\begin{equation*}
Y_{3}\leq4|\tau|\varepsilon_{1}\parallel Du\parallel_{L^{6}(g)}^{2}\parallel
u^{\prime}\parallel_{L^{6}(g)}
\end{equation*}
Elementary calculus gives 
\begin{equation*}
\parallel Du\parallel_{L^{6}(g)}\leq e^{-\frac{2}{3}\lambda_{m}}\parallel
Du\parallel_{6}\leq C_{0}|\tau|^{\frac{2}{3}}\parallel Du\parallel_{L^{6}}
\end{equation*}
and 
\begin{equation*}
\parallel u^{\prime}\parallel_{L^{6}(g)}\leq e^{\frac{1}{3}%
\lambda_{M}}\parallel u^{\prime}\parallel_{6}
\end{equation*}
The $L^{6\text{ }}$ norms can be estimated with $H_{1}$ norms using the
Sobolev inequality

\begin{equation*}
\parallel f\parallel_{6}\leq C_{\sigma}(\Vert f\Vert+\Vert Df\Vert)
\end{equation*}
applied to $f=u^{\prime}$ and $f=|Du|$ together with the inequality 
\begin{equation*}
D\mid f\mid\leq|Df|
\end{equation*}
we obtain 
\begin{equation*}
Y_{3}\leq4|\tau|\varepsilon_{1}C_{\sigma}e^{\frac{1}{3}(\lambda_{M}-%
\lambda_{m})-\lambda_{m}}[\Vert u^{\prime}\Vert+\Vert
Du^{\prime}\Vert][\Vert Du\Vert^{2}+\Vert D^{2}u\Vert^{2}]
\end{equation*}
hence, going back to the energies 
\begin{equation*}
Y_{3}\leq CC_{\sigma}|\tau|^{3}\varepsilon_{1}|[\varepsilon+\varepsilon
_{1}][\varepsilon^{2}+\varepsilon_{1}^{2}]
\end{equation*}

Finally

\begin{equation*}
Y_{4}\equiv|\int_{\Sigma_{t}}u^{\prime}\Delta_{g}N.\Delta_{g}u\mu_{g}|\leq|%
\tau|\varepsilon_{1}\parallel
u^{\prime}\parallel_{L^{4}(g)}\parallel\Delta_{g}N\parallel_{L^{4}(g)}
\end{equation*}
therefore, using laplacian and norms in conformal metrics and the previous
estimate of $\parallel u^{\prime}\parallel_{L^{4}(g)}$%
\begin{equation*}
Y_{4}\leq C_{\sigma}|\tau|^{\frac{3}{2}}e^{-2\lambda_{m}}e^{\frac{1}{2}%
\lambda_{M}}\varepsilon_{1}(\varepsilon^{2}+\varepsilon\varepsilon
_{1})\parallel\Delta N\parallel_{4}
\end{equation*}
The bound we have just computed of $\parallel D^{2}N\parallel_{4}$ gives
also a bound of $\parallel\Delta N\parallel_{4}$, hence 
\begin{equation*}
Y_{4}\leq|\tau|^{3}CC_{\sigma}\varepsilon_{1}(\varepsilon^{2}+\varepsilon
\varepsilon_{1})\varepsilon_{DN}(1+C(\varepsilon^{2}+\varepsilon
\varepsilon_{1})
\end{equation*}

Gathering the results gives the theorem.

\section{Decay of the total energy.}

We call \textbf{total energy} the quantity 
\begin{equation*}
E_{tot}(t)\equiv E(t)+\tau^{-2}E^{(1)}(t)\equiv\varepsilon^{2}+\varepsilon
_{1}^{2}
\end{equation*}
We define $y(t)$ to be the total corrected energy namely: 
\begin{equation*}
y(t)\equiv E_{1/4}(t)+\tau^{-2}E_{1/4}^{(1)}
\end{equation*}
\bigskip We have 
\begin{equation*}
E_{tot}(t)\leq\frac{1}{1-a_{t}}y(t)
\end{equation*}
with on each $\Sigma_{t}$ 
\begin{equation}
a_{t}\equiv\frac{|\tau|e^{\lambda_{M}}}{4\Lambda_{t}^{\frac{1}{2}}},\text{ \
\ }\Lambda_{t}\equiv\Lambda_{\sigma_{t}}
\end{equation}
The inequalities obtained for the corrected energies imply, with $\tau
=-t^{-1}$%
\begin{equation}
\frac{dy}{dt}=\frac{1}{t}[-y+A+B]
\end{equation}
where $A$ and $B$ are bounded by polynomials in $\varepsilon$ and $%
\varepsilon_{1}$ with terms of degree at least 3.

\begin{lemma}
Suppose that on ($\Sigma,\sigma)$ there is $\delta_{\sigma}>0$ such that the
first positive eigenvalue $\Lambda_{\sigma}$ is 
\begin{equation*}
(4\Lambda_{\sigma})^{-\frac{1}{2}}=\frac{1-\delta_{\sigma}}{\sqrt{2}}.
\end{equation*}
then if the energies are such that 
\begin{equation*}
CC_{\sigma}(\varepsilon^{2}+\varepsilon\varepsilon_{1})\leq\frac
{\delta_{\sigma}}{2}
\end{equation*}
then 
\begin{equation*}
1-a_{\sigma}\geq\frac{\delta_{\sigma}}{2}
\end{equation*}
The numbers $C$ and $C_{\sigma}$ are known numbers depending respectively on
the number $c$ of the hypothesis $H_{c}$ and on the metric $\sigma.$
\end{lemma}

Proof. By the definition of $a\equiv a_{\sigma}$ it holds that 
\begin{equation*}
1-a_{\sigma}=1-\frac{|\tau|e^{\lambda_{M}}}{\sqrt{2}}+\frac{\delta_{\sigma
}|\tau|e^{\lambda_{m}}}{\sqrt{2}}
\end{equation*}
which gives using the lower bound of $\lambda$ and the lemma 3 of the
section 8 ''conformal factor estimates'' 
\begin{equation*}
1-a_{\sigma}\geq\delta_{\sigma}-CC_{\sigma}(\varepsilon^{2}+\varepsilon
\varepsilon_{1})
\end{equation*}
from which the result follows.

\textbf{Hypothesis H}$_{\sigma}$ : 1. The numbers $C_{\sigma}$ are uniformly
bounded by a constant $M$ for all $t\geq t_{0}$ for which they exist.

2. There exists a constant $\delta>0$ such that the numbers $%
\Lambda_{\sigma} $, the first positive eingenvalues of $-\Delta_{\sigma_{t}}$
for functions with mean value zero, are such that 
\begin{equation*}
(4\Lambda_{\sigma})^{-\frac{1}{2}}=\frac{1-\delta_{\sigma}}{\sqrt{2}},\text{
\ \ with \ \ }\delta_{\sigma}\geq\delta.
\end{equation*}

\textbf{Hypothesis H}$_{E}$. The energies $\varepsilon_{t}^{2}$ and $%
\varepsilon_{1,t}^{2}$ satisfy as long as they exist an inequality of the
form 
\begin{equation*}
C(c,M)(\varepsilon^{2}+\varepsilon\varepsilon_{1})\leq\frac{\delta}{2}
\end{equation*}
where $C$ is a number depending only on the numbers $c$ and $M.$

We will prove the following theorem.

\begin{theorem}
Under the hypothesis H$_{c},H_{E}$ and H$_{\sigma}$ there exists a number $%
\eta$ such that if the total energy is bounded at time $t_{0}$ by $\eta$
then it satisfies at time $t=-$ $\tau^{-1}\geq t_{0}>0$ an inequality of the
form 
\begin{equation*}
tE_{tot}(t)\equiv t(\varepsilon^{2}+\varepsilon_{1}^{2})\leq
M_{tot}E_{tot}(t_{0})
\end{equation*}
where $M_{tot}$ depends only on $\delta.$
\end{theorem}

Proof. Under the hypothesis we have made the polynomials $A$ and $B$ are
bounded by polynomials in $y^{\frac{1}{2}}$ with terms of degree at least 3
and bounded coefficients depending only on $c,M,\delta.$

Take $\eta$ such that $y_{0}\equiv y(t_{0})<1.$ Then all powers of $y_{0}$
greater than 3/2 are less than $y_{0}^{3/2}$ and there exists a constant $%
M_{1},$ depending only on $c$, $\delta$ and $M$ such that 
\begin{equation*}
(A+B)_{t=t_{0}}\leq M_{1}y_{0}^{3/2}
\end{equation*}
Take $\eta$ such that moreover 
\begin{equation*}
y_{0}^{1/2}<\frac{1}{M_{1}}\text{ }
\end{equation*}
hence \ \ \ $\frac{dy}{dt}(t_{0})<0$ and $y$ starts decreasing, therefore
continues to satisfy $y<1.$ Therefore 
\begin{equation*}
A+B\leq M_{1}y^{3/2}
\end{equation*}
and $y$ satisfies the differential inequality 
\begin{equation}
\frac{dy}{dt}\leq-\frac{1}{t}(y-M_{1}y^{3/2})
\end{equation}
with always 
\begin{equation*}
y-M_{1}y^{3/2}>0
\end{equation*}
and, consequently, the differential inequality 
\begin{equation*}
\frac{dy}{y(1-M_{1}y^{1/2})}+\frac{dt}{t}\leq0
\end{equation*}
equivalently 
\begin{equation*}
\frac{dz}{z(1-M_{1}z)}+\frac{dt}{2t}\leq0,\text{ \ \ with \ \ }y=z^{2}
\end{equation*}
which gives by integration 
\begin{equation*}
\log\{\frac{z(1-M_{1}z_{0})}{(1-M_{1}z)z_{0}}\}+\log(\frac{t}{t_{0}})^{\frac{%
1}{2}}\leq0
\end{equation*}
that is 
\begin{equation*}
\frac{t^{\frac{1}{2}}z(1-M_{1}z_{0})}{(1-M_{1}z)t_{0}^{\frac{1}{2}}z_{o}}%
\leq1
\end{equation*}
in other words 
\begin{equation*}
t^{1/2}z+M_{1}z\frac{t_{0}^{1/2}z_{0}}{1-M_{1}z_{0}}\leq\frac{%
t_{0}^{1/2}z_{0}}{1-M_{1}z_{0}}
\end{equation*}
a fortiori 
\begin{equation*}
ty\leq\frac{t_{0}y_{0}}{(1-M_{1}z_{0})^{2}}.
\end{equation*}
We suppose for instance 
\begin{equation*}
z_{0}\leq\frac{1}{2M_{1}},
\end{equation*}
then 
\begin{equation*}
ty\leq4t_{0}y_{0}
\end{equation*}
Recall that under the H$_{c},$ H$_{E}$ and H$_{\sigma}$ hypotheses 
\begin{equation*}
E_{tot}(t)\leq\frac{1}{1-a_{t}}y(t)\leq\frac{2}{\delta_{t}}y(t),
\end{equation*}
also 
\begin{equation*}
y_{0}\leq\frac{1}{1-a_{0}}y_{0}\leq\frac{2}{\delta_{0}}E_{tot}(t_{0}),
\end{equation*}
The inequality for $y$ implies therefore 
\begin{equation*}
tE_{tot}(t)\leq M_{t}E_{tot}(t_{0})
\end{equation*}
\bigskip with, as announced, $M_{t}$ uniformly bounded: 
\begin{equation*}
M_{t}=\frac{4t_{0}}{(1-a_{t})1-a_{0})}\leq\frac{16t_{0}}{\delta_{t}%
\delta_{t_{0}}}\leq\frac{16t_{0}}{\delta^{2}}
\end{equation*}

\section{Teichmuller parameters.}

\subsection{Dirichlet energy.}

Let s and $\sigma$ be two given metrics on $\Sigma$ and $\Phi$ be a mapping
from $\Sigma$ into $\Sigma.$ The energy of the mapping $\Phi:(\Sigma
,\sigma)\rightarrow(\Sigma,s)$ is by definition the positive quantity: 
\begin{equation*}
E(\sigma,\Phi)\equiv\int_{\Sigma}\sigma^{ab}\frac{\partial\Phi^{A}}{\partial
x^{a}}\frac{\partial\Phi^{B}}{\partial x^{b}}s_{AB}(\Phi)\mu_{\sigma}
\end{equation*}
Consider the metric s as fixed. Elementary calculus shows that the energy $%
E(\sigma,\Phi)$ is invariant under a diffeomorphism $f$ of $\Sigma$ in the
following sense 
\begin{equation*}
E(\sigma,\Phi)=E(f_{\ast}\sigma,\Phi\circ f)
\end{equation*}
In the case where s and $\sigma$ both have negative curvature it has been
proved by Eells and Sampson that there exists one and only one harmonic map $%
\Phi_{\sigma}:(\Sigma,\sigma)\rightarrow(\Sigma,s)$ which is a
diffeomorphism homotopic to the identity, i.e. $\Phi_{\sigma}\in\mathcal{D}%
_{0}$. Such a harmonic map is equivariant under diffeomorphisms homotopic to
the identity, i.e. 
\begin{equation*}
\Phi_{f_{\ast}\sigma}=\Phi_{\sigma}\circ f,\text{ with }f\in\mathcal{D}_{0}
\end{equation*}
One is then led to the definition:

\begin{definition}
Given a metric s $\in M_{-1}$the Dirichlet energy $D(\sigma)$ of the metric $%
\sigma\in M_{-1}$ is the energy of the harmonic map $\Phi_{\sigma}\in%
\mathcal{D}_{0}$: 
\begin{equation*}
D(\sigma)\equiv E(\sigma,\Phi_{\sigma})
\end{equation*}
\bigskip It depends on the choice of the fixed metric s, but is invariant
under the action of diffeomorphisms included in $\mathcal{D}_{0}$ hence
defines a positive functional on the Teichmuller space $T_{eich}\equiv
M_{-1}/\mathcal{D}_{0}$.
\end{definition}

\begin{remark}
The energy of the mapping $\Phi:(\Sigma,\sigma)\rightarrow(\Sigma,s)$ as
well as the harmonic map $\Phi_{\sigma}$ are also invariant under conformal
rescalings of $\sigma.$ They can be used on the space of riemannian metrics
of negative curvature before the rescaling which restricts them to metrics
of curvature $-1$.
\end{remark}

The importance of the Dirichlet energy rests on the following theorem which
says that if $D(\sigma)$ remains in a bounded set of $R$ then the
equivalence class of $\sigma$ remains in a bounded set of $T_{eich}.$

\begin{theorem}
(Eells and Sampson) The Dirichlet energy is a proper function on Teichmuller
space.
\end{theorem}

\subsection{Estimate of the Dirichlet energy.}

We will require of the metric $\sigma_{t}$ that it remains, when t varies,
in some cross section of $M_{-1}$ (space of $C^{\infty}$ metrics with scalar
curvature -1) over the Teichmuller space, diffeomorphic to $R^{6G-6},$ $G$
the genus of $\Sigma$.

\textbf{Remark}. Following Andersson-Moncrief one can choose the cross
section as follows, having given some metric $s\in M_{-1}$. To an arbitrary
metric $\zeta$ $\in M_{-1\text{ }}$ we associate another such metric by its
pull back through $\Phi_{\zeta}^{-1}$%
\begin{equation*}
\psi(\zeta)=(\Phi_{\zeta}^{-1})_{\ast}\zeta
\end{equation*}
For any $f\in\mathcal{D}_{0}$ we have 
\begin{equation*}
\psi(f_{\ast}\zeta)=(\Phi_{f_{\ast}\zeta}^{-1})_{\ast}f_{\ast}\zeta=\psi
(\zeta)
\end{equation*}
hence the metric $\psi$ depends only on the equivalence class $Q$ of $\zeta$
through $\mathcal{D}_{0}.$ Thus one gets a cross section of $M_{-1}$ over
Teichmuller space, $Q\in T_{eich}\mapsto\psi(Q)\in M_{-1}$. If $Q$ remains
in a bounded set of $T_{eich\text{ }}$ then $\psi(Q)$ remains in a bounded
set of $M_{-1}$ i.e. all these metrics are uniformly equivalent.

We will estimate the Dirichlet energy $D(\sigma)\equiv E(\sigma,\Phi_{\sigma
}).$ We have, with $g_{ab}=e^{2\lambda}\sigma_{ab}$ 
\begin{equation*}
E(\sigma,\Phi_{\sigma})\equiv\int_{\Sigma}\sigma^{ab}\partial_{a}\Phi_{%
\sigma
}^{A}\partial_{b}\Phi_{\sigma}^{B}s_{AB}(\Phi_{\sigma})\mu_{\sigma}=\int_{%
\Sigma}g^{ab}\partial_{a}\Phi_{\sigma}^{A}\partial_{b}\Phi_{%
\sigma}^{B}s_{AB}(\Phi_{\sigma})\mu_{g}\equiv E(g,\Phi_{\sigma})\text{ }
\end{equation*}
If $\Phi_{\sigma}$ is a harmonic map from $(\Sigma,\sigma)$ into $(\Sigma,s)$
it is an extremal of the mapping 
\begin{equation*}
\Phi\rightarrow E(\sigma,\Phi)
\end{equation*}
and also an extremal of the mapping 
\begin{equation*}
\Phi\rightarrow E(g,\Phi)
\end{equation*}
We have, with $\frac{\partial E}{\partial g\text{ }}$ and $\frac{\partial E}{%
\partial\Phi}$ denoting functional derivatives (linear maps acting
respectively on $\frac{dg}{dt}$ and $\frac{d\Phi}{dt})$%
\begin{equation*}
\frac{d}{dt}E(g,\Phi)=\frac{\partial E}{dg}.\frac{dg}{dt}+\frac{\partial E}{%
\partial\Phi}.\frac{d\Phi}{dt}
\end{equation*}
We compute this derivative at a point ($\sigma,\Phi_{\sigma});$ we have , by
the extremality of $\Phi_{\sigma}$, ($\frac{\partial E}{\partial\Phi}%
)(\sigma,\Phi_{\sigma})=0$ . Therefore 
\begin{equation*}
\frac{d}{dt}D(\sigma)\equiv\{\frac{d}{dt}E(g,\Phi)\}_{(g,\Phi_{\sigma})}=\{%
\frac{\partial E}{\partial g}.\frac{dg}{dt}\}_{(g,\Phi_{\sigma})}
\end{equation*}
which gives using previous notations and the vanishing of the integral of a
divergence on a compact manifold 
\begin{equation*}
\frac{d}{dt}D(\sigma)=\int_{\Sigma_{t}}\{\overset{\_}{\partial}%
_{0}g^{ab}\partial_{a}\Phi_{\sigma}^{A}\partial_{b}\Phi_{\sigma}^{B}-N\tau
g^{ab}\partial_{a}\Phi_{\sigma}^{A}\partial_{b}\Phi_{\sigma}^{B}\}s_{AB}(%
\Phi_{\sigma})\mu_{g}
\end{equation*}
Recall that 
\begin{equation*}
\overset{\_}{\partial}_{0}g^{ab}=2Ng^{ac}g^{bd}k_{cd}=2Ne^{-4%
\lambda}h^{ab}+Ne^{-2\lambda}h^{ab}\tau
\end{equation*}
hence 
\begin{equation*}
\frac{d}{dt}D(\sigma)=\int_{\Sigma_{t}}2Ne^{-2\lambda}h^{ab}\partial_{a}%
\Phi_{\sigma}^{A}\partial_{b}\Phi_{\sigma}^{B}s_{AB}(\Phi_{\sigma})\mu
_{\sigma}
\end{equation*}
Using $0<N\leq2$ and e$^{-2\lambda}\leq\frac{\tau^{2}}{2}$ we find 
\begin{equation*}
|\frac{d}{dt}D(\sigma)|\leq2\tau^{2}\parallel h\parallel_{\infty}D(\sigma)
\end{equation*}
The bound of $\parallel h\parallel_{\infty}$ found in the section on $h$
estimates gives: 
\begin{equation*}
|\frac{d}{dt}D(\sigma)|\leq|\tau|CC_{\sigma}[\varepsilon+(\varepsilon
+\varepsilon_{1})^{2}]D(\sigma)
\end{equation*}

We recall the following lemmas.

\begin{lemma}
There exists an open subset $\Omega$ of $T_{eich}$ such that if the
equivalence class of $\sigma$ is in $\Omega$ and $\sigma$ is in a smooth
cross section of $T_{eich}$, then there exists a number $\delta>0$ such that 
$\Lambda(\sigma)\geq\frac{1}{8}+\delta$ and all constants $C_{\sigma}$ are
bounded by a fixed number $M.$
\end{lemma}

\begin{lemma}
There exists an interval I$\equiv$(a,b) of $R$ such that if the Dirichlet
energy (taken with some metric s) $D(\sigma)\in I$ then $\sigma$ projects
into $\Omega$. More precisely, there exists $\sigma_{0}$ projecting in $%
\Omega$ and given $\sigma_{0}$ there exists a number D such that if 
\TEXTsymbol{\vert}D($\sigma)-D(\sigma_{0})|\leq D$ then the hypothesis H$%
_{\sigma}$ is satisfied.
\end{lemma}

We will prove the following theorem

\begin{theorem}
Under the hypothesis H$_{c}$, H$_{E}$ and H$_{\sigma}$ there exists a number
M$_{D}$ depending only on the bounds in these hypothesis such that the
Dirichlet energy satisfies the inequality 
\begin{equation*}
|D(\sigma_{t})-D(\sigma_{0})|\leq M_{D}x_{0}^{\frac{1}{2}}\text{ \ \ with \
\ }x_{0}\equiv E_{tot}(t_{0})
\end{equation*}
\end{theorem}

Proof. Under the hypothesis that we have made the Dirichlet energy satisfies
the differential inequality (we have set $\tau=-t^{-1})$%
\begin{equation*}
|\frac{d}{dt}D(\sigma)|\leq D(\sigma)CM\{\frac{t^{\frac{1}{2}}[\varepsilon
+(\varepsilon+\varepsilon_{1})^{2}]}{t^{\frac{3}{2}}})
\end{equation*}
We recall the decay found for the total energy 
\begin{equation*}
tE_{tot}(t)\equiv t(\varepsilon^{2}+\varepsilon_{1}^{2})\leq
M_{tot}E_{tot}(t_{0})
\end{equation*}
with 
\begin{equation*}
M_{tot}\leq\frac{16t_{0}}{\delta^{2}}
\end{equation*}
\bigskip We have, using $t\geq t_{0}$ and $(\varepsilon+\varepsilon_{1})^{2}%
\leq2E_{tot}$ 
\begin{equation*}
t^{\frac{1}{2}}(\varepsilon+(\varepsilon+\varepsilon_{1})^{2})\leq t^{\frac
{1}{2}}E_{tot}^{\frac{1}{2}}(t)+2t_{0}^{-\frac{1}{2}}tE_{tot}(t)
\end{equation*}
Using the decay of the total energy (section 11) and the assumption $%
x_{0}\equiv E_{tot}(t_{0})<1$ we find that there exists a number $M_{2}$
depending only on $c,M$ and $\delta$ such that 
\begin{equation*}
|\frac{d}{dt}D(\sigma)|\leq D(\sigma)\frac{M_{2}x_{0}{}^{\frac{1}{2}}}{t^{%
\frac{3}{2}}}
\end{equation*}
We deduce from this inequality, by elementary calculus, abbreviating $%
D(\sigma)$ to $D$ and $D(\sigma_{0})$ to $D_{0},$%
\begin{equation*}
\frac{d}{dt}|D-D_{0}|\leq|\frac{d}{dt}(D-D_{0})|\leq\lbrack|D-D_{0}|+D_{0}]%
\frac{M_{2}x_{0}^{\frac{1}{2}}}{t^{\frac{3}{2}}}
\end{equation*}
By the Gromwall lemma $|D-D_{0}|$ is for $t\geq t_{0}$ bounded by the
solution of the associated differential equality with initial value zero,
which gives 
\begin{equation*}
|D-D_{0}|\leq\lbrack D_{0}M_{2}x_{0}^{\frac{1}{2}}\int_{t_{0}}^{t}t^{-\frac
{3}{2}}dt]\exp(M_{2}x_{0}^{\frac{1}{2}}\int_{t_{0}}^{t}t^{-\frac{3}{2}}dt)
\end{equation*}
hence, as announced 
\begin{equation*}
|D_{\sigma_{t}}-D_{\sigma_{0}}|\leq M_{D}x_{0}^{\frac{1}{2}}
\end{equation*}
with (recall that $x_{0}\leq1$) 
\begin{equation*}
M_{D}=D_{\sigma_{0}}2M_{2}t_{0}^{-\frac{1}{2}}\exp(2M_{2}t_{0}^{-\frac{1}{2}%
})
\end{equation*}

\section{Global existence.}

\begin{theorem}
Let ($\sigma_{0},q_{0})\in C^{\infty}(\Sigma_{0})$ and ($u_{0},\overset{.}{u}%
_{0})\in H_{2}(\Sigma_{0},\sigma_{0})\times H_{1}(\Sigma_{0},\sigma_{0})$ be
initial data for the polarized Einstein equations with U(1) isometry group
on the initial manifold $M_{0}\equiv\Sigma_{0}\times U(1)$ ; suppose that $%
\sigma_{0}$ is such that $R(\sigma_{0})=-1$ and the first positive
eigenvalue $\Lambda_{0}$ of -$\Delta_{\sigma_{0}}$ (for functions with mean
value zero) is such that 
\begin{equation*}
\Lambda_{0}>\frac{1}{8}.
\end{equation*}
Then there exists a number $\eta>0$ such that if 
\begin{equation*}
E_{tot}(t_{0})<\eta
\end{equation*}
these Einstein equations have a solution on $M\times\lbrack t_{0},\infty),$
with initial values determined by $\sigma_{0},q_{0},u_{0},\overset{.}{u}%
_{0}. $ The orthogonal trajectories to the space sections $M\times\{t\}$
have an infinite proper length.
\end{theorem}

Proof. It results from the local existence theorem that we only have to
prove that $E_{tot}(t)$ does not blow up. We have in the previous sections
made the following hypothesis, to hold for all $t\geq t_{0}$ for which the
involved quantities exist

Hypothesis H$_{c}$. There exists a number $c>c_{0}=\varepsilon_{v_{0}}>0$
such that

1. 
\begin{equation*}
\varepsilon_{v}\leq c
\end{equation*}

2. 
\begin{equation*}
\varepsilon_{0}\leq\frac{1}{2(1+2ce^{2c})}
\end{equation*}

Hypothesis H$_{D}.$ The Dirichlet energy is such that 
\begin{equation*}
|D(\sigma)-D(\sigma_{0})|\leq d
\end{equation*}
where $d>0$ is a given number such that the above inequality implies the
hypothesis $H_{\sigma}.$

Hypothesis $H_{E}.$ The total energy is such that 
\begin{equation*}
E_{tot}(t)\leq c_{E}
\end{equation*}
where $c_{E}$ is a number depending only on $c$ and $d$.

Under these hypothesis we have obtained the following result: there are
numbers $A_{i}$ depending only on $c$ and $d$ such that 
\begin{equation*}
\varepsilon_{v}\leq A_{1}E_{tot}(t_{0})
\end{equation*}
and 
\begin{equation*}
tE_{tot}(t)\leq A_{2}E_{tot}(t_{0})
\end{equation*}
and 
\begin{equation*}
|D(\sigma)-D(\sigma_{0})|\leq A_{3}E_{tot}^{\frac{1}{2}}(t_{0})
\end{equation*}
Now consider the triple of numbers 
\begin{equation*}
\{X_{t}\equiv\varepsilon_{v_{t}},x_{t}\equiv E_{tot}(t),Z_{t}\equiv
|D(\sigma_{t})-D(\sigma_{0})|\}
\end{equation*}
We have shown that the hypothesis 
\begin{equation*}
X_{t}\leq c,\text{ \ }x_{t}\leq c_{E},\text{ }Z_{t}\leq d
\end{equation*}
and smallness conditions on $x_{0},$ imply the existence of numbers $A_{i}$
depending only on $c$, $c_{E}$ and $d$ such that 
\begin{equation*}
X_{t}\leq A_{1}x_{0},\text{ \ \ }tx_{t}\leq A_{2}x_{0},\text{ \ \ }Z_{t}\leq
A_{3}x_{0}^{\frac{1}{2}}
\end{equation*}
Therefore there exists $\eta>0$ such that $x_{0}\leq\eta$ implies that the
triple belongs to the subset $U_{1}\subset R^{3}$ defined by the
inequalities: 
\begin{equation*}
U_{1}\equiv\{X_{t}<c,\text{ \ }x_{t}<c_{E},\text{ }Z_{t}<d\}
\end{equation*}
For such an $\eta$ the triple either belongs to $U_{1}$ or to the subset $%
U_{2}$ \ defined by 
\begin{equation*}
U_{2}\equiv\{X_{t}>c\text{ \ \ or }x_{t}>c_{E}\text{ \ \ or \ }Z_{t}>d\}
\end{equation*}
These subsets are disjoint. We have supposed that for $\ t=t_{0}$ it holds
that $(X_{0},x_{0},Z_{0})\in U_{1}$ hence, by continuity in t, $%
(X_{t},x_{t},Z_{t})\in U_{1}$ for all t. We have proved the required a
priori bounds.

The orthogonal trajectories to the space sections $M\times\{t\}$ have an
infinite proper length since the lapse $N$ is bounded below by a strictly
positive number.

$^{1}$ Supported in part for the NSF contract n$%
{{}^\circ}%
$ PHY-9732629 to Yale university

Aknowledgements. We thank L.\ Andersson for suggesting the use of corrected
energies. We thank the University Paris VI, the ITP in Santa Barbara, the
University of the Aegean in Samos and the IHES in Bures for their
hospitality during our collaboration.

\textbf{References.}

V.\ Moncrief Reduction of Einstein equations for vacuum spacetimes with U(1)
spacelike isometry group, Annals of Physics 167 (1986), 118-142

A.\ Fisher and A.\ Tromba Teichmuller spaces, Math. Ann. 267 (1984) 311-345.
Cf. also Y.\ Choquet-Bruhat and C.\ DeWitt-Morette Analysis Manifolds and
Physics, Part II, supplemented edition North Holland, 2000.

J.\ Cameron and V.\ Moncrief The reduction of Einstein's vacuum equations on
space times with U(1) isometry group. Contemporary Mathematics 132 (1992)
143-169.

D.\ Christodoulou and S.\ Klainerman The global non linear stability of the
Minkowsi space, Princeton University Press 1992.

Y.\ Choquet-Bruhat and V.\ Moncrief Existence theorem for solutions of
Einstein equations with 1 parameter spacelike isometry group, Proc. Symposia
in Pure Math, 59, 1994, H.\ Brezis and I.E.\ Segal ed. 67-80

Y.\ Choquet-Bruhat and J.\ W.\ York Well posed system for the Einstein
equations, C. R. Acad. Sci. Paris 321, (1995) 1089-1095

L.\ Andersson, V.\ Moncrief and A. Tromba On the global evolution problem in
2+1 gravity J.\ Geom. Phys. 23 1997 n$%
{{}^\circ}%
3-4$,1991-205

\end{document}